\documentstyle[aps,prd,preprint,tighten]{revtex}
\begin{document}
\preprint{FERMILAB-Pub-93/317-A}
\overfullrule=0pt

\draft
\author{Andrew H. Jaffe\cite{byline}}
\address{Enrico Fermi Institute and Department of Astronomy \& Astrophysics\\
         5640 S. Ellis Avenue, Chicago, IL 60637-1433\\
	 NASA/Fermilab Astrophysics Center, Fermi National Accelerator
	 Laboratory,\\ Batavia, IL 60510-0500}
\title{Quasi-Linear Evolution of Compensated Cosmological Perturbations:
       The Nonlinear Sigma Model}
\date{\today}
\maketitle
\begin{abstract}

We consider the evolution of perturbations to a flat FRW
universe that arise from a ``stiff source,'' such as a
self-ordering cosmic field that forms in a global
symmetry-breaking phase transition and evolves via the Kibble
mechanism. Although the linear response of the normal matter to
the source depends on the details of the source dynamics, we
show that the higher-order non-linear perturbative equations
reduce to a form identical to those of source-free Newtonian
gravity in the small wavelength limit.  Consequently, the
resulting $n$-point correlation functions and their spectral
counterparts will have a hierarchical contribution arising from
this gravitational evolution (as in the source-free case) in
addition to that possibly coming from non-Gaussian initial
conditions.  We apply this formalism to the $O(N)$ nonlinear
sigma model at large $N$ and find that observable differences
from the case of initially Gaussian perturbations and Newtonian
gravity in the bispectrum and higher-order correlations are not
expected on scales smaller than about $100\,h^{-1}{\rm Mpc}$.

\end{abstract}

\section{Introduction}

This paper attempts to combine several disparate streams of work
in the study of cosmological perturbations.  Since the early
work of Lifschitz\cite{LL}, Sachs and Wolfe\cite{SW} and others,
the evolution of large scale matter perturbations to a spatially
homogeneous and isotropic expanding universe has been
well-studied.  More recently, this work has been refined to
apply to a more complicated universe containing both radiation
(with equation of state $p/\rho=1/3$) and non-relativistic
matter ($p=0$) constituents, using a gauge-invariant
approach\cite{bardeen,ks,fmb}.  In addition, large $N$-body
simulations have been used to examine the fully nonlinear
problem of gravitational evolution on scales much smaller than
the Hubble radius, where the Newtonian limit of general
relativity is sufficient\cite{nbody}. On the other hand, certain
aspects of the full non-linear evolution equations do remain
amenable to a more analytic approach.  An area of work that has
received considerable attention in the past decade is the
examination of higher-order corrections to the linear Newtonian
equations of motion for the matter in the universe within the
framework of perturbation
theory\cite{lss,fry84,goroff,bernardeau}.  These higher-order
corrections are intimately related to the evolution of the
multi-point correlations of the mass (and therefore, of the
galaxies), such as the three-point function,
$\xi_3\propto
\langle\delta({\bf x}_1)\delta({\bf x}_2)\delta({\bf x}_3)\rangle$,
the skewness, or, in fourier-space, the bispectrum,
$B\propto\langle\delta({\bf k}_1)\delta({\bf k}_2)\delta({\bf
k}_3)\rangle$.  As increasingly large galaxy surveys become
available, observational information on these moments of the
galaxy distribution has been extended to scales sufficiently
large that this perturbative approach is expected to be
reliable.  In the case of an initially Gaussian density field
for which all reduced higher-order correlations vanish, this
``quasi-nonlinear'' gravitational evolution leads to a scaling
hierarchy of correlation functions:
$\bar\xi_n\propto\bar\xi_2^{n-1}$, where $\bar\xi_n$ is the
volume-averaged $n$-point correlation function, and the constant
of proportionality depends weakly on the power spectrum or
two-point function of the perturbations. Even with an initially
non-Gaussian distribution, it may be possible to distinguish the
primordial component from that due to gravitational evolution on
large scales.

The standard cold dark matter (CDM) model for structure
formation invokes a mechanism such as an early epoch of
inflation\cite{generalrefs} to generate primordial adiabatic
density fluctuation. Inflation results in a spatially flat
universe overall; quantum fluctuation of the inflaton field lead
to an initially scale-invariant spectrum of density
perturbations (with details possibly depending upon the specific
model of inflation) which evolve thereafter under the influence
of gravity. In the simplest models of inflation, this initial
distribution of perturbations is Gaussian, so the hierarchical
results noted above obtain for the resulting correlation
functions. The specific mechanism which drives inflation is
inextricably linked to the fundamental particle physics model of
the universe. Of course, in addition to providing a mechanism
for the generation of perturbations, inflation also has the
advantage of solving the horizon and flatness problems which are
otherwise left unexplained.

In another class of theories, topological defects (or some other
classical field configuration) such as cosmic strings, domain
walls or textures act as a continual source for the creation of
density perturbations\cite{generalrefs}.  These structures are
the remnants of a symmetry-breaking phase transition of a cosmic
field and are thus, like inflation, a cosmological relic of the
underlying particle physics.  In these scenarios, the field is
initially laid down randomly before the phase transition, on
scales larger than the Hubble radius at that time. Once the
symmetry is broken, the field tries to align itslef in order to
minimize its energy density. However, this alignment can only
occur coherently on scales where the field has come into causal
contact with itself---within the Hubble volume. As the universe
expands, the Hubble volume increases and the field orders itself
on larger and larger scales---the Kibble mechanism\cite{kibble}.
If the symmetries of the field and its initially random
configuration require it, topological defects may result. For
example, a broken discrete symmetry like ${\cal Z}_n$ produces
domain walls, a broken $O(2)$ or $U(1)$ symmetry can result in
(gauge or global) cosmic strings, $O(3)$ in monopoles, and
global $O(4)$ in global texture. The breaking of a global $O(N)$
symmetry with $N>4$ does not lead to topological defects, but
does result in spatial field gradients and consequently to
perturbations to the energy density.  Unlike the inflationary
scenario, the Kibble mechanism quite generically produces
density fluctuations with an initially non-Gaussian
distribution.  (Of course, these theories do not provide a
natural solution to the horizon and flatness problems and in
this case it is usually assumed that the universe begins in
a perfectly homogeneous and isotropic state with $\Omega=1$
before the symmetry-breaking phase transition. Thus, in these
models one
relegates the solution of these puzzles to initial
conditions, or to an earlier inflationary epoch driven by a
field too weakly-coupled to produce the density
perturbations responsible for large-scale structure.)

Within the class of scenarios that create structure via the
Kibble mechanism, there is a further conceptual split between
global and local (or gauge) symmetries. In a local theory, the
large-scale gradient energy of the field is compensated by the
gauge field so all the field energy is concentrated in localized
defects.  Thus, only true topologically stabilized
configurations can result. In global theories, on the other
hand, non-topological configurations (textures) can result that are
nonetheless long-lived because they require energy to ``unwind''
a configuration by forcing it off its vacuum manifold. In
large-$N$ models, these field configurations persist simply
because causality constrains them to align only on scales
smaller than the Hubble radius, so field gradients persist
for approximately a Hubble time. (Also, a
new class known as ``semi-local'' defects has been studied, in
which a gauge theory admits defects which are stabilized by the
dynamics of theory, not the underlying topology of the symmetry
groups\cite{semilocal}.)

In this paper, we shall modify and extend aspects of a formalism
that has been developed by Veeraraghavan and Stebbins \cite{vs}
to study the perturbations due to a ``stiff source'' such as
these cosmic field configurations formed by the Kibble
mechanism.  A ``stiff source'' evolves in the homogeneous and
isotropic background metric of the universe; the back-reaction
of the metric perturbations onto the source is considered to be
negligible. We shall, however, explicitly account for
compensation: the initial response of the matter fields to the
stress-energy of the stiff source.  We shall extend previous work to
allow the perturbations of the matter and radiation fluids in
the universe to enter the quasi-nonlinear regime and examine the
modifications to the perturbative solution to the equations of
motion that result.  This, in turn, allows us to study the
resulting correlation functions and may modify the scaling
hierarchy in such scenarios.

Using this formalism, we shall concentrate on a specific class
of models in which the Kibble mechanism is responsible for the
initial generation of density perturbations, the nonlinear
$O(N)$ sigma model\cite{turoksigma,davis,penetal}. These models
arise as the low-energy limit of the breaking of a global $O(N)$
symmetry to $O(N-1)$.  Although the analytic calculations we
shall perform rely on the large-$N$ limit, in which there are
only spatial gradients, but no topological defects, we can also
extract some information about the behavior of this theory for
small $N$ where defects play a role.

The plan of this paper is as follows. In Section
\ref{sec:formal}, we develop the equations of motion for the
stiff source in the background metric and of the matter fields in the
perturbed universe, using the longitudinal gauge, and develop a
perturbative expansion about homogeneity and isotropy. In Section
\ref{sec:sigma}, we apply this formalism to the nonlinear sigma model, and
examine the higher-order correlations of the matter. Finally, we
present our conclusions. In an appendix, we show some useful
results for the distribution of the fields in an $O(N)$ model.

\section{Compensated Perturbations}
\label{sec:formal}

\subsection{Perturbations in the Longitudinal Gauge}

We will consider linear metric fluctuations about a homogeneous
FRW background.  We work in comoving conformal coordinates with
metric signature $(+,-,-,-)$ and assume a spatially flat background metric of
\begin{equation}
ds^2 = a^2\eta_{\mu\nu}dx^\mu dx^\nu
     = a^2\left(d\eta^2 - \delta_{ij}dx^idx^j\right),
\end{equation}
where $\eta_{\mu\nu}$ is the Minkowski metric.  Here, $\eta$ is
conformal time, related to proper time by $dt=a\,d\eta$.
Throughout, we shall use a prime to denote the derivative with
respect to conformal time, and define a conformal expansion rate
${\cal H}=a'/a$. The unperturbed Einstein equations in the flat
($\Omega=1$) Freedman-Robertson-Walker (FRW) universe with
mean background density $\rho$ and pressure $p$ and vanishing
cosmological constant ($\Lambda=0$) are
\begin{equation}
{\cal H}^2={8\pi G\over3}a^2\rho;\qquad
{a''\over a}={4\pi G\over3}a^2(\rho-3p).
\end{equation}
In a matter-dominated universe ($p=0$), the scale factor
$a\propto\eta^2$ and in a radiation-dominated universe
($p=\rho/3$), $a\propto\eta$.  The Hubble constant is
$H_0=100h\,{\rm km}\:{\rm sec}^{-1}\:{\rm Mpc}^{-1}$ and the present
Hubble radius is $H_0^{-1}=3000\,h^{-1}{\rm Mpc}$. We normalize the
conformal time by setting $\eta_0=2H_0^{-1}=6000\,h^{-1}{\rm Mpc}$
today.  We will write the perturbed metric as
\begin{equation}
g_{\mu\nu} = a^2(\eta_{\mu\nu} + h_{\mu\nu}).
\end{equation}
In the longitudinal gauge
($h_{0i} = 0$) to first order in $h$,
$g^{\mu\nu} = a^{-2}(\eta_{\mu\nu} - h_{\mu\nu})$ or
$g^{00} = a^2(1-h_{00})$ and $g^{ij} = -a^{-2}(\delta_{ij} + h_{ij})$.

In the usual longitudinal (often called conformal-Newtonian)
gauge analysis, only scalar perturbations are considered, in
which case the metric perturbations are determined by two scalar
variables (``potentials''), $h_{00}=2\phi$ and $h =
\delta_{ij}h_{ij} = 6\psi$.  In this paper, we will in addition
allow nonzero vector and tensor perturbations, for which
${\widetilde h}_{ij} = h_{ij} - \delta_{ij}h/3$ is nonzero. Most analyses
of density perturbations in an expanding universe have used the
synchronous gauge ($h_{00}=h_{0i}=0$) \cite{lss} (although a
gauge-invariant approach has recently become popular
\cite{bardeen,ks,fmb}).  We have chosen the longitudinal gauge
in order to more easily compare our results with the
perturbative Newtonian equations of motion (and thereby gain
insight into higher order correlations which have previously
been studied only in the Newtonian limit); the price of this is
retaining a nonzero component $h_{00}$. More importantly, in the
longitudinal gauge we can consider situations in which the
density perturbation amplitude $\delta\equiv\delta\rho/\rho$ is
large, while the metric perturbations are still small.

A few more words about our approximation scheme are in order.
Because we shall only compute the Einstein tensor to first order
in the metric perturbation, the perturbed Einstein equations
$\delta G_{\mu\nu} = 8\pi G\delta T_{\mu\nu}$ (where here
$\delta T_{\mu\nu}$ represents the perturbation to all the
stress-energy in the universe, including matter, radiation and
any sources such as those discussed below) can only be
considered first order --- {\it in the metric perturbation}.
Thus, we must ignore terms like $h_{\mu\nu}\cdot\delta
T_{\alpha\beta}$ for consistency.  However, this does not
determine the form that the stress-energy must take: it may
contain $v^2$ or $v\delta$ terms. (Below, we shall calculate the
four-velocity to order $v^2$.)  Thus, these equations can be
valid for ``large'' values of the density perturbation amplitude
$\delta\sim1$ as long as we still have $v^2, h_{\mu\nu}\ll1$.
Note that this is a gauge-dependent statement. In fact,
performing the same exercise in a comoving synchronous gauge, we
find that $h'=2\delta'$ to all orders in the matter perturbation
variables for pressureless, nonrelativistic matter, so this
scheme would not be successful---we can only use synchronous
coordinates until streamlines of the matter flow intersect and
caustics form.  The advantage of the synchronous formalism is
that one can define a comoving gauge where the nonrelativistic
matter component has zero velocity for all time ({\it i.e.}, as
long as the coordinate construction is consistent).  In the
longitudinal gauge we are considering here we still have to
calculate velocities. However, the gravitational potential
(metric perturbation) in this gauge is in general suppressed
compared to the density perturbation by the square of the size
of the perturbation relative to the Hubble radius, as in the
usual Newtonian Poisson equation:
$h\sim\phi\sim(\lambda/H^{-1})^2\delta$ for perturbations of
scale $\lambda$. (Of course, this equation is itself only
applicable for scales smaller than the Hubble radius, beyond
which relativistic corrections become important as in {\it
e.g.}, Eq.\ (\ref{cdmsoln}) below.)

On scales smaller than the Hubble radius, the metric
perturbation remains small even when the matter perturbation
becomes nonlinear.  In particular, we expect this approximation
to be valid as long as $h^2\ll\delta$; with first order
quantities and the Poisson equation to relate the potential to
the density perturbation, this is equivalent to $\left(\lambda/
H^{-1}\right)^4\ll{1/\delta}.$ As expected, our approximation
will continue to hold on scales much smaller than the horizon as
long as $\delta\lesssim1$.  Conversely, for sufficiently small
scales the linear metric perturbation approximation holds for
even larger values of the density perturbation.  As long as the
average source stress-energy is negligible compared to the fluid
components and as long as we continue to ignore backreaction,
this approximation holds in the same regimes as the normal
Newtonian limit.

In the longitudinal gauge, ${\widetilde h}_{ij}$ does not
contain a scalar part as it does in the synchronous gauge, so in
addition to being manifestly traceless, it satisfies
$\partial_i\partial_j{\widetilde h}_{ij}=0$.
Although the matter variables become more complicated when we
allow higher-order terms to enter the equations, the metric
perturbation is still usefully decomposed into geometrically
distinct parts ({\it i.e.}, scalar, vector and tensor)\cite{ks}.
Normally, only the scalar part of the peculiar velocity, which
can be expressed as the gradient of some velocity potential,
appears in the equations of motion for other scalar quantities
like the density perturbation $\delta$.  With higher order
terms, however, products like $v^i\delta$ or $v^iv^j$ decompose
as a vector and symmetric tensor, respectively, with scalar,
vector, and (for $v^iv^j$) tensor parts that are not simply
related to the scalar and vector parts of the original $v^i$.
Thus, for example, $\delta$, although a scalar, can appear in
higher-order equations containing the vector perturbations to
the metric, and the vector part of $v^i$ can occur in the scalar
equations of motion ({\it e.g.}, Eq.\ (\ref{G00}) which contains
$v^2=v^iv^i$).

As in \cite{vs} (who instead calculated in synchronous gauge),
we assume a universe filled with a multi-component perfect-fluid
stress-energy $T^\mu{}_{\!\nu}=\sum_{n} T_{(n)}^\mu{}_{\!\nu}$,
where $n$ labels the fluid component, as well
as a ``stiff source'' with stress-energy $\Theta_{\mu\nu}$. The
source makes a negligible contribution to the spatially-averaged
stress-energy and is covariantly conserved with respect to the
{\it background} metric $a^2\eta_{\mu\nu}$:
\begin{equation}\label{stiffeq}
  \Theta'_{00} + {\cal H}\left(\Theta_{00} +
  \Theta\right) = \partial_i\Theta_{0i};
  \qquad   \left(a^2\Theta_{0i}\right)' = a^2\partial_j\Theta_{ij}.
\end{equation}
Here $\Theta\equiv\Theta_{ii}$ is the spatial trace of the source
stress energy.

Again, we do not consider the back-reaction on the stiff source
of the metric perturbations.  In principle, we could calculate
it and its effects in an iterative, perturbative fashion,
writing the full source stress-energy as a sum of a part
propagating in the FRW background as in Eq.\ (\ref{stiffeq}) and
another part propagating in the full perturbed metric. This
perturbation would, in turn, be fed back into the field
equations for the metric perturbation. This would result in new
terms of order
$(\partial_\mu h_{\alpha\beta})\Theta_{\gamma\delta}$, still at
linear order in the metric. Unless either the source stresses or
the metric perturbation are large, these terms should be
negligible in comparison to those involving the linear fluid
perturbations.

The bulk of the universe is filled with several perfect
fluids with individual stress-energies
\begin{equation}
  T_{(n)}^\mu{}_{\!\nu} =
    (\rho_n + \delta\rho_n + p_n+\delta p_n)u_n^\mu u_{n\nu} -
      (p_n+\delta p_n)\delta^\mu{}_{\!\nu},
\end{equation}
where $n$ labels the fluid component, $\rho_n$ and $p_n$ refer
to the background density and pressure of each fluid, and
$\delta\rho_n$, $\delta p_n$ to their respective perturbations.
The four-velocity of each fluid is
$u^\nu={\bar u}^\nu+\delta u^\nu$,
normalized to $u^\nu u_\nu=+1$ with this signature. In
the expanding background, the fluid components are at rest, with
four-velocity (suppressing the fluid label $n$)
\begin{equation}
        {\bar u}^\mu = (a^{-1},0,0,0),\qquad {\bar u}_\nu = (a,0,0,0)
\end{equation}
and the velocity perturbations are
\[
      \delta u^0 = {1\over 2} a^{-1}(v^2-h_{00}),\qquad
      \delta u_0 = {1\over 2} a(v^2 + h_{00}),
\]
\begin{equation}
      \delta u^i = -a^{-2} \delta u_i =
                   a^{-1}v^i\left(1+{1\over 2} v^2\right)
\end{equation}
This defines the peculiar velocity 3-vector, $v^i=u^i/u^0$.
We have kept terms of order $v^2$, and ignored terms of
order $v\cdot h$, for the consistency of our approximation scheme.
(Note that the form of the $v^2$ terms is exactly as one would expect
from the usual special-relativistic four velocity expanded to
this order.)

\subsection{Equations of Motion for the Matter and Metric Fields}

We will assume that each of the perfect fluid components is
separately conserved. Except when there is significant energy
transfer between the components ({\it e.g.}, at the epoch of
recombination) this is an excellent approximation.
To first order in the metric perturbation, the equations of
motion for a single fluid component, $\nabla_{\!\mu}
T_{(n)}^\mu{}_{\!\nu}=0$, become
\begin{mathletters}\label{conservation}\begin{eqnarray}\label{T0i}
 \delta' &+&v^2 \partial_0\left[\delta(1+\zeta)\right]
 + \left(1+w+\delta(1+\zeta)\right) \partial_0 v^2
 +\nabla\cdot\left[\left(1+w+\delta(1+\zeta)\right) {\bf v}\right] \nonumber\\
 &+&3{\cal H}\left[v^2\left(1+w+\delta(1+\zeta)\right)(w+{1\over3})
     + \delta(\zeta-w) +
 v^2(1+w)(w-c^2)\right]\nonumber\\
 &=&3\psi'(1+w)
\end{eqnarray}\noindent
and
\begin{eqnarray}\label{Tij}
 {\cal H} v^i\left[\left(1+w+\delta(1+\zeta)\right)(1-3w) +
     3(1+w)(w-c^2)\right] +
 \left[\delta(1+\zeta)\right]'v^i &+& \nonumber\\
 \left(1+w+\delta(1+\zeta)\right) v^{i\prime} +
    \partial_j\left[\left(1+w+\delta(1+\zeta)\right) v^i v^j\right]
 + \partial_i(\zeta\delta)
 &=&-(1+w)\partial_i\phi
\end{eqnarray}\end{mathletters}\noindent
where $\delta=\delta\rho/\rho$, $\zeta=\delta p/\delta\rho$, the
fluid sound speed $c^2=\partial p/\partial\rho$, $w=p/\rho$, and we
have suppressed the subscript $n$ on all fluid quantities. To
linear order, $\zeta_n=c_n^2$ and we shall often take
$w_n=\zeta_n=c_n^2$ as well. In the background, the fluid
density evolves according to the zeroth-order equations of
motion, $\rho_n'=-3(1+w_n)\rho_n$.  Notice that these equations
do not depend explicitly upon the stiff source.

The perturbed Einstein equations $\delta G_{\mu\nu} = 8\pi
G(\sum_{(n)}\delta T_{(n)\mu\nu}+\Theta_{\mu\nu})$ become
\begin{mathletters}\label{einstein}\begin{eqnarray} \label{G00}
 4\pi G \tau_{00} &=& \nabla^2\psi \nonumber\\
 &=& 3{\cal H}\left(\psi'+{\cal H}\phi\right) + 4\pi G
 a^2\sum_n\rho_n\left[\delta_n +
 v_n^2\left(1+w_n+\delta_n(1+\zeta_n)\right)\right] + 4\pi G\Theta_{00}\\
 4\pi G \tau_{0i} &=& \nabla\psi' -
      {1\over4}\partial_j{\widetilde h}'_{ij} \nonumber\\
 &=&-{\cal H}\nabla\phi-4\pi G a^2\sum_n\rho_n
     {\bf v}_n\left(1+w_n+\delta_n(1+\zeta_n)\right) +
 4\pi G\Theta_{0i}
  \label{G0i}\\
 4\pi G \tau_{ij} &=& \delta_{ij}\left[\psi''+
  {1\over 2}\nabla^2(\phi-\psi)\right]
 -{1\over 2}\partial_i\partial_j(\phi-\psi) +
  {1\over 2}{}^{(3)}{\widetilde G}_{ij}
 -{1\over 4} {\widetilde h}_{ij}''\nonumber\\
 &=& {1\over 2}{\cal H}{\widetilde h}'_{ij}-
 \delta_{ij}\left\{\left[2{a''\over a}-{\cal H}^2\right]\phi
 +{\cal H}(2\psi'+\phi')\right\} \nonumber\\ &+& 4\pi G a^2
 \sum_n\rho_n\left[\delta_n\zeta_n\delta_{ij}+
       \left(1+w_n+\delta_n(1+\zeta_n)\right)v_n^iv_n^j\right]
 +4\pi G\Theta_{ij}   \label{Gij}
\end{eqnarray}\end{mathletters}\noindent
where
\begin{equation}
{}^{(3)}{\widetilde G}_{ij} = {1\over 2}\left[
  \nabla^2{\widetilde h}_{ij}-\partial_j\partial_k{\widetilde h}_{ik}-
  \partial_i\partial_k{\widetilde h}_{jk}\right]
\end{equation}
is the trace-free part of the spatial Einstein tensor in this gauge.
We have written the equations such that the form of the
stress-energy pseudotensor $\tau_{\alpha\beta}$
is manifest. By Eqs.\ (\ref{einstein}), the pseudotensor
manifestly obeys
\begin{equation}\label{taumunu}
 \partial_\mu \tau^{\mu\nu} = 0; \qquad
 \tau^{\mu\nu}=\eta^{\alpha\mu}\eta^{\beta\nu}\tau_{\alpha\beta}.
\end{equation}
In effect, the pseudotensor can be written as
\begin{equation}
\tau_{\mu\nu} = T_{\mu\nu}+\Theta_{\mu\nu}+t^{\rm grav}_{\mu\nu}
\end{equation}
where $T_{\mu\nu}$ and $\Theta_{\mu\nu}$ are the fluid and
source stress-energies, respectively, and the remaining term,
$t^{\rm grav}_{\mu\nu}$ represents the ``stress-energy of the
gravitational field.'' As we see in Eqs.\ (\ref{einstein}), this
term contains those parts of the Einstein tensor that are either
nonlinear in the metric perturbation (although these terms do
not appear in our approximation) or those that would vanish for
a constant scale factor $a'=0$.

The pseudotensor is conserved with respect to a Minkowski
background; it thereby embodies the concept of a conserved
stress-energy for the {\em entire system} including the
``gravitational energy'' as well as the matter and source
stress-energy \cite{vs,weinberg,mtw}, the combined system
propagating on a flat, Minkowski background. However, no
individual fluid or gravitational components of the pseudotensor
can be singled out as being separately conserved in this
Minkowski space; moreover, its definition is gauge-dependent.
It is useful for setting up the conditions of compensation on
scales larger than the Hubble radius: it can describe the flow
of energy between matter and ``gravity'' which only occurs
causally, on scales within the Hubble volume. Note also that to
this order, the metric perturbation only appears in terms
containing factors of ${\cal H}$---in a true Minkowski background
({\it i.e.}, not expanding), the pseudotensor only differs from the
``real'' stress-energy tensor at second and higher orders in the
metric perturbation\cite{weinberg,mtw}.

The space-space component of the Einstein equations may be
rewritten as a trace and traceless part:
\begin{mathletters}\label{Gspace}\begin{eqnarray}
 \left[2{a''\over a} - {\cal H}^2\right]\phi
 &+& {\cal H}(2\psi'+\phi') + \psi'' +
{1\over 3}\nabla^2(\phi-\psi)\nonumber\\
 &=& 4\pi G a^2\sum_n\rho_n\left[\delta_n\zeta_n +
     {1\over3}\left(1+w_n+\delta_n(1+\zeta_n)\right)v_n^2\right]
    +{4\pi G\over 3}\Theta
\label{Gtrace}\\
 \left[{1\over 3}\delta_{ij}\nabla^2-\partial_i\partial_j\right](\phi &-& \psi)
 + {1\over 2}\left[\nabla^2{\widetilde h}_{ij}-
  \partial_j\partial_k{\widetilde h}_{ik}-
  \partial_i\partial_k{\widetilde h}_{jk}
 -{\widetilde h}''_{ij}-2{\cal H}{\widetilde h}'_{ij}\right]\nonumber\\
 &=& 8\pi G
  a^2\sum_n\rho_n\left(1+w_n+\delta_n(1+\zeta_n)\right)\left(v_n^iv_n^j-
     {1\over 3}\delta_{ij} v_n^2\right)
    +8\pi G {\widetilde\Theta}_{ij}. \label{Gtracefree}
\end{eqnarray}\end{mathletters}\noindent
Here, Eq.\ (\ref{Gtrace}) is the spatial trace of Eq.\
(\ref{Gij}) ($4\pi G\tau_{kk}$) and Eq.\ (\ref{Gtracefree})
comes from subtracting off that trace from Eq.\ (\ref{Gij})
($\tau_{ij}-\delta_{ij}\tau_{kk}/3$).  Also,
${\widetilde\Theta}_{ij} = \Theta_{ij}-\delta_{ij}\Theta/3$ is
the spatial trace-free part of the source stress energy.

(Of course, the conservation Eqs.\ (\ref{conservation}) are not
independent of the Einstein equations, which imply that the {\it
sum} of the individual fluid stress energies is covariantly
conserved via the Bianchi identity. In an approximation that we will
consider several times later, a universe with only one fluid in
addition to the stiff source, the conservation equations are
redundant and can be derived from the Einstein equations by
using the conservation formulas for the
pseudostress-energy (Eq.\ (\ref{taumunu})) and eliminating the source terms
$\Theta_{\mu\nu}$ by using their background equations of
motion, Eqs.\ (\ref{stiffeq}).)

Obviously, these equations (\ref{conservation}, \ref{einstein},
\ref{Gspace}) cannot be solved analytically for a general
equation of state, with general initial data and an arbitrary
stiff source.  However, we can explore various limits and
approximations that are useful in different theories of
structure formation.

\subsection{Correspondence with Newtonian Equations of Motion}
\label{sec:Newton}

If we consider a universe filled only with pressureless matter
($w=c^2=\zeta=0$) and no stiff source ($\Theta_{\mu\nu}=0$) we
have the conditions of the standard matter-dominated universe.
If we further assume the limit of scales small compared to the
Hubble radius and small velocities, $v\ll c$, then we have the
situation usually described by the Newtonian limit.  In the
equations of motion for the matter (Eqs.\ (\ref{conservation}))
and the Einstein Eqs.\ (\ref{einstein},\ref{Gspace}), we drop
terms like $\partial_0v^2$ in this limit but we keep terms like
$\partial_iv^2$ because we assume that the matter component may
vary over small scales. Note that we do not assume that the
density perturbation $\delta$ is small, so we do not ignore
terms that contain $\delta$ in spatial derivatives, even when
multiplied by small quantities like ${\cal H}^2$ or ${\bf v}$.  In this
case, the spatial Einstein Eqs.\ (\ref{Gspace}) only describe
the evolution of the scalar components of the metric, $\phi$ and
$\psi$; the tensor-mode equations drop out as expected as these
components evolve completely independently. The
$i\neq j$ Einstein Eq.\ (\ref{Gtracefree}) then requires
$\phi=\psi$.  The trace of the $i-j$ Eq.\ (\ref{Gtrace})
gives
\begin{equation}
 \left(2{a''\over a} - {\cal H}^2\right)\phi +3{\cal H}\phi'+\phi''=0
\end{equation}
with solution $\phi={\rm const}$ (plus a decaying mode
proprotional to $\eta^{-5}$).  With these results, the $0-0$
Einstein Eq.\ (\ref{G00}) and the stress-energy conservation
Eqs.\ (\ref{conservation}) (which are not independent from the
Einstein equations) become
\begin{mathletters}\label{newton}\begin{eqnarray}
 {3\over2}{\cal H}^2\delta &=& \nabla^2\phi, \\
 \delta' + \nabla\cdot\left[(1+\delta){\bf v}\right] &=& 0, \\
 {\cal H}{\bf v} + {\bf v}' + \left({\bf v}\cdot\nabla\right){\bf v}
 &=& -\nabla\phi,
\end{eqnarray}\end{mathletters}\noindent
where we have used the second equation to eliminate $\delta'$ in
the final equation.

These equations are just those that come from a purely Newtonian
analysis of density perturbations in the fluid limit: the
Poisson equation, the continuity equation, and the Euler
equation. It should not be surprising that this approximation
scheme has yielded these results: in Newtonian theory, the
potential $\phi\sim v^2$, so it is suppressed to second order
compared to the matter variables when the velocities are small,
and ignoring terms like $\phi\delta$ and $\phi v$ should be
sufficient to this level of approximation. This approximation
can be made more precise by expanding separately in two small
parameters: the size of the metric perturbation $h$ and the
ratio of the length scale of the perturbation in question
to the Hubble length. This expansion generates the Newtonian and
post-Newtonian approximations in a cosmological
setting\cite{weinberg,postnewton}.

If we consider a simple perturbation expansion in both the
matter velocity ${\bf v}$ and density perturbations $\delta$ (with
both quantities the same order), and an initially Gaussian
density field ({\it e.g.}, from inflation), these equations have been
shown to produce a scaling hierarchy of correlation functions in
the quasi-nonlinear regime\cite{lss,fry84,bernardeau}:
$\bar\xi_n\propto\bar\xi_2^{n-1}$, where $\bar\xi_n$ is the
volume-averaged $n$-point correlation function or moment of the
distribution ({\it e.g.}, $\bar\xi_3$ is the skewness). For the
unaveraged multi-point correlation functions
$\xi_n({\bf x}_1,\ldots,{\bf x}_n)$ or their fourier-space spectral
counterparts $P_n({\bf k}_1,\ldots,{\bf k}_n)$, the $n$-point
functions are proportional to sums of appropriate symmetric products of
$(n-1)$ two-point functions.
For the simple case of the bispectrum,
$B_{123}=Q(P_1P_2+P_1P_3+P_2P_3)$ as in Eq.\ (\ref{qk}) below;
the trispectrum is a product of three power spectra,
$T_{1234}=R_a[P(k_1)P(|{\bf k}_1+{\bf k}_2|)P(k_3) + {\rm sym.}]
+R_b[P(k_1)P(k_2)P(k_3) + {\rm sym.}],$ where ``sym.''
represents the further terms of the same form necessary to keep
$T$ symmetric in its arguments. Here, $Q$, $R_a$, $R_b$ and
their generalizations to higher $n$-point functions are
constants that depend on the configuration of the ${\bf k}_i$ and
possibly on the initial power spectrum.

For this hierarchical result to apply, two additional
assumptions are necessary: the decaying mode component of the
solution for the density perturbation must become negligible,
and the velocity-field must be curl-free, so it can be expressed
as the gradient of some potential; any leftover vortical
component of the velocity decays away proportional to $a^{-1}$,
so this assumption is justified at late times.

It is interesting to examine how these equations are modified
even in a universe with only a matter component when
relativistic effects are included. That is, we no longer make
the Newtonian approximation of small velocities and length
scales that was used to derive Eqs.\ (\ref{newton}). Even to
{\em linear} order, these equations have corrections for
perturbation wavelengths larger than the Hubble radius
($\lambda\gg H^{-1}$, which we shall hereafter refer to as {\em
superhorizon} scales).  The Poisson equation simply becomes the
non-relativistic matter ($w=0$) version of Eq.\ (\ref{G00}) with
$\phi=\psi$, and the continuity equation gains a potential term
(relaxing the assumption $\phi={\rm const}$):
\begin{mathletters}\begin{eqnarray}
 {3\over2}{\cal H}^2\delta &=& \nabla^2\phi-3{\cal H}\phi'-3{\cal H}^2\phi\\
 3\phi' &=& \delta' + \nabla\cdot\left[(1+\delta){\bf v}\right]
\end{eqnarray}\end{mathletters}\noindent
while the Euler equation is unchanged. The linear solution to
these equations is well known, and is presented in the following
section, Eqs.\ (\ref{cdmsoln}).
The full nonlinear equations also have corrections of order
$v^2$. However, we expect that when these terms are significant,
our approximation that the metric perturbations are still of
linear order will break down.

\subsection{Full Equations in the Linear Regime}

One useful application of Eqs. (\ref{conservation}-\ref{Gspace})
is to a universe consisting only of pressureless matter,
radiation and, possibly, a stiff source, in the linear regime
(where now we are considering perturbations linear in
everything: $\delta$, $v^i$ and $h_{\mu\nu}$).  Because these
equations are linear, the full power of previously developed
techniques can be applied. In particular, we can use the
geometrical decomposition of these equations to study the
scalar, vector, and tensor modes separately. Density
fluctuations in particular only couple to the scalar mode of the
metric perturbation to this order, $\phi$ and $\psi$, as well as
the scalar parts of the peculiar velocity and, if present, the
stiff source. In a universe with non-relativistic matter ($w=c^2=\zeta=0$)
and radiation ($w=c^2=\zeta=1/3$) fluids, the linear Einstein Eqs.\
(\ref{G00},\ref{G0i},\ref{Gspace}) for the scalar modes become
\begin{mathletters}\begin{eqnarray}
\nabla^2\psi - 3{\cal H}\left(\psi'+{\cal H}\phi\right) -4\pi G \Theta_{00} &=&
            {3\over2}{\cal H}^2(\Omega_m\delta_m+\Omega_r\delta_r) \\
\nabla^2\left(\psi' + {\cal H}\phi\right) - 4\pi G \partial_i\Theta_{0i} &=&
           -{3\over2}{\cal H}^2(\Omega_m\nabla\cdot{\bf v}_m+
	           {4\over3}\Omega_r\nabla\cdot{\bf v}_r) \\
{\cal H}(2\psi'+\phi') + \psi'' + {1\over 3}\nabla^2(\phi-\psi) &=&
           {1\over 2}{\cal H}^2\Omega_r\delta_r+{4\pi G \over 3}\Theta\\
\left[{1\over 3}\delta_{ij}\nabla^2-\partial_i\partial_j\right](\phi-\psi) =
           8\pi G \widetilde\Theta^{(s)}_{ij} &=& -8\pi G
           \left[{1\over 3}\delta_{ij}\nabla^2-\partial_i\partial_j\right]s
\end{eqnarray}\end{mathletters}\noindent
where $\Omega_m$ and $\Omega_r$, respectively, represent the
ratio of the matter and radiation densities to the
(time-dependent) critical density, with $\Omega_m+\Omega_r=1$ by
assumption.  $\widetilde\Theta^{(s)}_{ij}$ is the scalar,
trace-free part of $\Theta_{ij}$, and can be written as above in
terms of a second-order linear operator acting on a unique
scalar function $s(\eta, {\bf x})$, determined, for example, by
expanding $\Theta_{ij}$ in terms of appropriate harmonic
functions. (In particular,
$\Theta_{ij}\approx\delta_{ij}\Theta/3$ on superhorizon scales,
where the differential operator that projects out the $s$
component approximately vanishes, as do all spatial derivatives.)
Thus, the final equation tells us that $\phi-\psi = -8\pi G s$
up to a spatial constant (and in a universe with no stiff
source, $\phi=\psi$ as usual).  Even without a stiff source,
these equations cannot be solved exactly in the presence of both
matter and radiation fluids except in various limits.  In a
matter-dominated universe ($\Omega_m=1$), the final two
equations give $\phi$ and $\psi$ in terms of the source, and the
first two in turn give the density and velocity.

Because we are again considering linear perturbation theory, the
complete solution in the matter-dominated era is just given by
the sum of the usual pure matter solution (vanishing $\Omega_r$ and
$\Theta_{\mu\nu}$) and a ``particular'' solution generated by the
$\Theta_{\mu\nu}$ terms. The source-free matter solution is:
\begin{mathletters}\label{cdmsoln}\begin{eqnarray}
\phi_c &=& \psi_c = C_1({\bf x}) + C_2({\bf x})\eta^{-5} \approx {\rm const}\\
\delta_c &=& {1\over 6}\left[\eta^2\nabla^2C_1 - 12C_1 +
             \left(\eta^2\nabla^2C_2 + 18C_2\right)\eta^{-5}\right]\nonumber\\
  &\approx& {1\over 6} \left(\eta^2\nabla^2\phi_c - 12\phi_c\right)
\end{eqnarray}\end{mathletters}\noindent
Superhorizon scales can be defined by the condition that
$\eta\nabla\rightarrow0$; in fourier space where
$\nabla\rightarrow k\sim1/\lambda$, this becomes the appropriate
$k\eta\rightarrow0$, since we also have $\eta\sim H^{-1}$.  On
these scales, $\delta\approx-2\phi\approx{\rm const}$, so
perturbations do not grow, while on small scales the usual
Poisson equation applies.  In this case, the equations can be
rewritten as the linearized version of the Newtonian case
considered above in Section \ref{sec:Newton}.

In the presence of a source term, we must add the appropriate
particular solution to this homogeneous solution.
After setting $\phi-\psi=-8\pi G s$ as discussed above, the
solution for the potential $\phi$ is
\begin{equation}
\phi = \phi_c + {1\over 5}\int \chi\eta\, d\eta -
        {1\over 5}\eta^{-5}\int\eta^6\chi\,d\eta; \qquad
\chi(\eta, {\bf x}) = 8\pi G\left(
        {1\over 6}\Theta - 2{\cal H} s' -s'' + {1\over 3} \nabla^2s \right).
\end{equation}
Given these solutions for the scalar potentials, the density
fluctuation can be computed immediately. In the presence of a
source term, the question of initial conditions for these
fluctuations is crucial. We wish to express the fact that, at
the time of ``source creation'' (e.g., a phase transition) the
universe is initially homogeneous, and the matter variables can
only respond causally to the source stress-energy---that is, the
initial anisotropies of the stiff source stresses must be
compensated by the matter fields. Within the Hubble volume, this
produces actual perturbations in the fluid component; over
superhorizon scales, the matter fields cannot vary coherently
due to causality, so the universe must have zero curvature
outside the Hubble volume.  This isocurvature nature of these
perturbations is embodied in the Minkowski-space conservation of
the stress-energy pseudotensor $\tau^{\mu\nu}$ whose components
were given above.  In the initially homogeneous universe, the
pseudoenergy $\tau_{00}\equiv{\cal E}=0$ as well as
$\Theta_{\mu\nu}=0$. As is manifest from the Einstein Eqs.\
(\ref{einstein}), the evolution equations for the pseudotensor
are given by
\begin{equation}\label{taueq}
  \tau'_{00} = \partial_i\tau_{0i};  \qquad  \tau'_{0i} = \partial_j\tau_{ij}.
\end{equation}

When the phase transition, a Poisson random process, occurs, we
expect both of these quantities $\Theta_{\mu\nu}$ and
$\tau_{\mu\nu}$ to be initially uncorrelated with themselves and
thus gain white noise power spectra on superhorizon scales.
That is, we Fourier decompose all quantities as
\begin{equation}
f({\bf k})=
\int d^3x\,f({\bf x})e^{i{\bf k}\cdot{\bf x}}
\end{equation}
and we shall write $f\sim k^n$; this means
that the power spectrum $|f_k|^2\propto k^{2n}$. Thus, we expect
$\Theta_{\mu\nu}\sim k^0$ and $\tau_{\mu\nu}\sim k^0$.
The $i-j$ components only occur in spatial derivatives, so these
white noise spectra should obtain for these components for all
time as long as $k\eta\ll 1$ ({\it i.e.}, while the modes are
outside the Hubble radius and out of causal contact with
themselves), where we are considering the contribution from
logarithmic intervals about a wavenumber $k$.  From the
evolution equations, we then see that $\tau_{0i}\sim k$ and
$\tau_{00}\sim k^2$, whereas $\Theta_{0i}\sim k$ and
$\Theta_{00}\sim k^0$ (since the $\Theta_{00}$ equation is
dominated by the ${\cal H}$ term). Thus, the quantities $\tau_{00}$,
$\tau_{0i}$, and $\Theta_{0i}$ should remain negligible for
$k\eta\ll 1$. A similar argument using the Einstein equations
reveals that the density and metric fluctuations should
initially have white noise spectra on superhorizon scales, but
that velocities should fall as ${\bf v}\sim k$.

Therefore, the superhorizon scale density perturbation is simply
given from the $0-0$ Einstein Eq.\ (\ref{G00}) by the relation
$\tau_{00} = 0$:
\begin{equation}
\sum_n\Omega_n\delta_n = -{2\over3}{\cal H}^{-2}\Theta_{00} -
  2{\cal H}^{-1}\psi' -2\phi
\end{equation}
Here, this equation holds for a multi-component universe; below,
we shall specialize to a universe dominated by a single
component with $p_n/\rho_n=w_n$ and
$\Omega_n\approx1$.
On large scales $\psi'\simeq\delta'_m/3(1+w_n)$ (from
Eq.\ (\ref{T0i}))
and
$\phi\simeq{\rm const}$.
For the primary fluid component, this gives the superhorizon evolution
equation for $\delta$,
\begin{equation}
\delta+{2\over3{\cal H}(1+w_n)}\delta'=
      -{2\over3}{\cal H}^{-2}\Theta_{00} -2\phi
\end{equation}
For a universe with scale factor
$a\propto\eta^\alpha$,
\begin{equation}
\delta(\eta,{\bf x}) = -{1+w\over\alpha}\eta^{-3\alpha(1+w)/2}\int d\eta\,
              \Theta_{00}(\eta,{\bf x}) \eta^{3\alpha(1+w)/2 + 1} -
	      2\phi(\eta,{\bf x})
\end{equation}
In a radiation-dominated universe ($w=1/3$) this gives
$\delta_r$.  (In order to calculate the superhorizon matter
perturbation in this case, we define the entropy perturbation
$\sigma=3\delta_r/4-\delta_m$. If as causality requires there is
no initial entropy perturbation on superhorizon scales,
$\delta_m=3\delta_r/4$ and it is unnecessary to explicitly worry
about the details of the several components.)

In particular, if the stiff source obeys a scaling relation
$\Theta_{00}\propto\eta^{-2}$ (see Section \ref{sec:sigma}), then
\begin{equation}
\delta = -{2\over3\alpha^2}\eta^2\Theta_{00} - 2\phi.
\end{equation}
Both of these terms are time-independent, so the perturbations
do not grow with time outside the Hubble radius.  The same
result was derived by Davis {\it et al.}\ in the synchronous
gauge\cite{davis}.  This leads to the usual scale-invariant
Harrison-Zel'dovich power spectrum: $\delta\rho/\rho\approx{\rm
const}$ at horizon-crossing.  In the nonlinear sigma model to be
discussed later, the matter-dominated era evolution does obey
$\Theta_{00}\propto\eta^{-2}$, but in the radiation era
$\Theta_{00}\propto\eta^{-2}\ln(\eta/\eta_1)$,
where $\eta_1$ is the conformal time of the symmetry-breaking
phase transition, so the evolution of the density perturbation
is modified by a (divergent) logarithmic term. Thus, the
initially white-noise spectrum (constant amplitude on all scales
at one time) has been transformed into a scale-invariant
spectrum (constant amplitude at horizon-crossing) by the
gravitational action of the stiff source.

Compensation thus insures that the pseudoenergy ${\cal E}$ vanishes
on superhorizon scales for all time. This fact in turn gives the
initial condition for the perturbation on a given scale at
Horizon crossing. (Note that this is not the case for the usual
primordial adiabatic perturbations discussed in CDM models,
where the initial perturbation spectrum is considered as a given
before the equations are integrated.)

In order to calculate the evolution of density fluctuations well
inside the horizon, it is easiest to use the first of Eqs.\
(\ref{taueq}), $\tau'_{00}=\partial_i\tau_{0i}$ and the Einstein
equations which define the components of $\tau_{\mu\nu}$, Eqs.\
(\ref{einstein}). Here, it is assumed that the universe contains
both radiation and matter fluids.  After some algebra to
eliminate the matter velocity and the potential $\psi$, this
component of Eq.\ (\ref{taueq}) reduces to
\begin{eqnarray}\label{pseudeq}
{\cal H}\delta_m'' + {\cal H}^2\delta_m'-{3\over2}{\cal H}^3\Omega_m\delta_m
&-&3{\cal H}^3(1-\Omega_m)\delta_r +
3{\cal H}^3(\Omega_m-2)\phi+3{\cal H}^2\phi'
               \nonumber\\
&=& 4\pi G\left(\partial_i\Theta_{0i}-\Theta'_{00}\right)
\end{eqnarray}
With the initial condition ${\cal E}=0$ due to compensation,
in a matter-dominated universe, this becomes simply
\begin{equation}\label{phisol}
{3\over2}{\cal H}^2(\delta + 2\phi) + {\cal H}\delta' =
4\pi G\int^\eta d\eta\,\partial_i\Theta_{0i}
\end{equation}
where we have ignored the contribution of $\Theta_{00}$ to the
total energy density inside the horizon. The solution is
\begin{equation}
\delta = {4\pi G\over2}\eta^{-3}
\int^\eta d\eta_1\, \eta_1^4\int^{\eta_1}d\eta_2\,
     \partial_i\Theta_{0i}(\eta_2)
-6\eta^{-3}\int^\eta d\eta\, \eta^2\phi + K\eta^{-3},
\end{equation}
where we have explicitly included a decaying mode
$\propto\eta^{-3}$.  A similar equation was derived in the
synchronous gauge by Davis {\it et al.}\cite{davis}.
In the synchronous gauge, the
term involving $\phi$ is not present and the equation can be
solved directly for the density perturbation. This term,
however, is exactly as one would expect in performing the gauge
transformation from synchronous gauge to longitudinal gauge for
a scalar quantity $\delta\rho$:
$\delta_l = \delta_s-3{\cal H}(1+w)a^{-1}\int d\eta\, a\phi$
\cite{fmb}, where $l$ and $s$ refer to quantities in the
longitudinal and synchronous gauge, respectively. In the
longitudinal gauge, unfortunately, things are more complicated
and we must use our solution for the potential from above
subject to the initial condition of ${\cal E}(k\eta\rightarrow0)=0$
for a complete solution. However, on extreme subhorizon scales,
we can in general neglect terms like ${\cal H}^2 \phi$ in the above
equations---the longitudinal and synchronous coordinates nearly
coincide. For example, in taking the Newtonian limit of the
$\delta G_{00}$ Einstein equation in a pure CDM universe, we
drop such a term in order to derive its nonrelativistic limit,
the Poisson equation.  Thus, the growing mode solution reduces
to solely the first term above. In this case, and also assuming
that the source has the power law form $\int
d\eta\,\partial_i\Theta_{0i}\propto\eta^\gamma$ (as in Section
\ref{sec:sigma} below, the growing mode
solution is
\begin{equation}\label{md_mode}
\delta = {4\pi G\over 2(5+\gamma)}\eta^2\int d\eta\,\partial_i\Theta_{0i}.
\end{equation}
Davis {\it et al.}\ derived this result for the specific case of
$\gamma=0$ in the synchronous gauge\cite{davis}, in which case the solution
is of the same form as the normal pure-CDM growing mode
$\delta\propto a\propto\eta^2$ (which would be the solution here in
the case of a nonzero initial value of the pseudoenergy:
$\delta\propto{\cal E}_0\eta^2$.)

In a radiation-dominated universe, the situation is more
complicated. Note that we are interested in the evolution of {\it
matter} perturbations in this situation. Thus, we take
$\Omega_m\approx0$ in Eq.\ (\ref{pseudeq}) above, giving
\begin{equation}
\delta_m'' + {\cal H} \delta_m' \approx 4\pi G
{\cal H}^{-1}(\partial_i\Theta_{0i}-\Theta'_{00})
\end{equation}
where the terms involving $\phi$ and $\delta_r$ have been
ignored well inside the horizon. The solution is
\begin{equation}
\delta = C_1 + C_2\ln\eta+ 4\pi G
\int^\eta d\eta_2\,\eta_2^{-1}\int^{\eta_2} d\eta_1\,\eta_1^2
(\partial_i\Theta_{0i}-\Theta'_{00})_{\eta_1},
\end{equation}
which include the usual constant and logarithmic terms, in addition to one
due to the source evolution. Again considering the case of
$\int d\eta\,\partial_i\Theta_{0i}\propto\eta^\gamma$, the density
perturbation due to the stiff source becomes
\begin{equation}\label{rd_mode}
\delta =
4\pi G {\gamma\over(2+\gamma)^2}\eta^2\int d\eta\,\partial_i\Theta_{0i}.
\end{equation}

In the particular case that
$\int d\eta\,\partial_i\Theta_{0i}\approx{\rm const}$,
($\gamma\approx0$) the response to the stiff source exactly
mimics the usual isocurvature perturbation scenario:
perturbations only grow inside the horizon and during
matter-domination, $\delta\propto a$ (or logarithmically
during radiation-domination).  If the source varies more
rapidly, then corrections to this behavior may become important.

\subsection{Higher Order Equations With a Stiff Source}
\label{sec:nlstiff}

On superhorizon scales, the linear equations considered in the
last section cannot be extended to higher orders in the matter
variables within the approximation of linear metric fluctuations.
Outside the horizon, large density fluctuation create large
fluctuations in the metric, and we would at least need to treat
higher orders in $h_{\mu\nu}$ as well as the matter variables
(if not solve the full equations numerically).  Well inside the
horizon, however, the situation is different and the density
fluctuation amplitude may be large for a small metric perturbation in the
longitudinal gauge. Thus, we may still approximate the initial
conditions as a vanishing pseudoenergy, but retain the nonlinear
terms on small scales.

Further, on small scales, we may again assume that the usual conditions
of the Newtonian limit apply: potentials and velocities are small
and slowly-varying. Of course, because of the presence of the source
terms, it is important to at least check that these conditions still
hold.  From the linear solution of Eq.\ (\ref{phisol}), we expect the
potential to vary rapidly in regions of spacetime where the source
(specifically, its scalar parts $\Theta$ and $s$) is rapidly-varying
itself. For the sigma-model discussed below, the time scale of variation
is always the Hubble time, so this assumption is justified in that
case. For models with topological defects, the regions around the
defects themselves would generally have large stresses, and we expect
this analysis to fail there.

 In order to compare with the usual Newtonian equations, the relevant
equations are again the covariant conservation equations and the $0-0$
Einstein equation, now supplemented by the traceless $i-j$ equation to
take into account the anisotropic stresses of the stiff stress-energy.
On subhorizon scales, these give
\begin{mathletters}\begin{eqnarray}\label{nsource}
 {3\over2}{\cal H}^2\delta &=& \nabla^2\psi, \\
 \delta' + \nabla\cdot\left[(1+\delta){\bf v}\right] &=& 0,
          \label{contwsource}\\
 {\cal H}{\bf v} + {\bf v}' + \left({\bf  v}\cdot\nabla\right){\bf v}
&=& -\nabla\phi, \\
 \phi - \psi &=& -8\pi G s.
\end{eqnarray}\end{mathletters}\noindent
(A rapidly-varying potential would add a term $3\psi'$ to the right hand
side of Eq.\ (\ref{contwsource}).)
The linear solution to these equations for stiff sources
originating in a smooth universe is just that derived above.
Although these equations do not explicitly contain the term
$\Theta_{0i}$ that appears in the above solution (Eq.\
(\ref{md_mode})), recall that the source necessarily obeys
conservation with respect to the background metric, so the two
forms cannot be independent; they are related by the equations
of motion for the source, Eqs.\ (\ref{stiffeq}).

Except for the presence of the $s$ term, these are exactly the
same equations as in the Newtonian matter-dominated case without
sources, discussed
above, Eqs.\ (\ref{newton}). Note also that $s$ occurs in a
``pure source'' term---it is a function that is supplied from
outside of this set of equations and the set of equations are
linear in that function.

This permits a particularly simple derivation of the
higher-order equations of motion---{\it i.e.}, the deviations from the
linear solution above. We can write the various quantities as
\begin{equation}
\delta = {}^{\scriptscriptstyle(1)}\!\delta +
         {}^{\scriptscriptstyle(2)}\!\delta + \cdots
\end{equation}
where the superscript represents an $n$-th order quantity
({\it i.e.}, $O[({}^{\scriptscriptstyle(1)}\!\delta)^n]$).  Once
the linear equations have been subtracted off, the resulting
equations give differential equations for the higher-order terms
in terms of lower-order ones. These equations will have exactly
the same structure as in the pure Newtonian case (Section
\ref{sec:Newton} above); the only dependence upon the stiff
source will be implicitly through the
${}^{\scriptscriptstyle(1)}\!\delta$ solution.  That is, we can
write the linear solution as a functional of the stiff source,
${}^{\scriptscriptstyle(1)}\!\delta={}^{\scriptscriptstyle(1)}\!\delta[s]$.
As usual in perturbation theory, the higher-order quantities
only explicitly depend on the lower order
${}^{\scriptscriptstyle(n)}\!\delta$:
\begin{equation}
{}^{\scriptscriptstyle(n)}\!\delta=
{}^{\scriptscriptstyle(n)}\!\delta
      \left[{}^{\scriptscriptstyle(1)}\!\delta,\ldots,
            {}^{\scriptscriptstyle(n-1)}\!\delta\right],
\end{equation}
where the dependence on the source is implicit in
${}^{\scriptscriptstyle(1)}\!\delta$, and thus in all
higher-order terms as well.

In the Newtonian case, it is usually assumed that the initial
(linear) fluctuations have a Gaussian distribution; in the
presence of a stiff source the distribution of the density
fluctuations depend crucially upon the distribution of the
source through the linear solution. However, the distribution of
a stiff source is not a general property, but depends upon the
specific model. As noted above, if $\int
d\eta\,\partial_i\Theta_{0i}\approx{\rm const}$, the matter
perturbation behaves just as it does under the usual Newtonian
evolution. Then, the problem is equivalent to the Newtonian one
with the additional possibility of non-Gaussian initial
conditions\cite{fry+scherrer}.

\section{The Nonlinear Sigma Model}
\label{sec:sigma}
\subsection{Background Evolution}

Although the derivation thus far has been completely general and valid
for any stiff source that obeys conservation with respect to the
background (except as noted), most sources do not have a simple analytic
form that can be ``plugged in'' to these equations. One notable
exception is the $O(N)$ nonlinear sigma model in the limit of large $N$,
which is exactly soluble in an expanding $\Omega=1$ FRW universe. In
order to be reasonably self-contained, we shall follow Davis {\it et al.}
\cite{davis} closely in this section (note however the difference in
normalizations of the random variables $\alpha_{\bf k}^a$ and the
definition of the power-law exponent $\alpha$).

First, consider the lagrangian for the $O(N)$ fields $\phi^a$,
$a=1\ldots N$:
\begin{equation}
{\cal L} = {1\over 2} \nabla_{\!\mu}\phi^a \nabla^\mu\phi^a -V(\phi).
\end{equation}
where we raise and lower indices with the background metric
$a^2\eta_{\mu\nu}$. If the $O(N)$ symmetry is broken to
$O(N-1)$, the potential will have a nonzero minimum,
$V(\phi^a_0)=0$; the usual example is the broken ``phi-fourth''
potential, $V(\phi)=\lambda(\phi^2 - \phi_0^2)^2$ with
$\phi^2=\phi^a\phi^a$. With this sort of potential, the $\phi^a$
will have $N$ massless modes corresponding to angular
excitations, and one massive mode corresponding to radial
excitations of the $N$-vector $\phi$. In the low-energy or
strong-coupling limit, we can represent the potential of the
massless modes with a lagrange multiplier term,
\begin{equation}
{\cal L} = {1\over 2} \nabla_{\!\mu}\phi^a \nabla^\mu\phi^a
 +{1\over 2}\lambda\left(\phi^2-\phi_0^2\right)
\end{equation}
Thus, the $N$ fields $\phi^a$ are fixed to the $(N-1)$-sphere
vacuum manifold: $\phi^a\phi^a=\phi_0^2$.  The fields begin
randomly distributed before the phase transition; as they come
inside the Hubble radius, they come into causal contact with one
another and order themselves to minimize the gradient energy
(the first term of the Lagrangian) on those scales.

The equation of motion is
\begin{equation}\label{sigmaeom}
\nabla^\mu\nabla_{\!\mu}\phi^a =
   {\phi^b\nabla^\mu\nabla_{\!\mu}\phi^b\over\phi_0^2}\phi^a =
  -{\nabla_{\!\mu}\phi^b\nabla^\mu\phi^b\over\phi_0^2}\phi^a
\end{equation}
where the second equality follows from the vacuum manifold
constraint enforced by the lagrange multiplier
\cite{spergeltexture} (and thus is only
strictly true for the nonlinear sigma model, but not in models
with a ``physical'' potential).
At large $N$, we can replace the bilinear
$\nabla_{\!\mu}\phi^b\nabla^\mu\phi^b$ with its (ensemble or
spatial) average and the scaling ansatz
\begin{equation}
\langle\nabla_{\!\mu}\phi^b\nabla^\mu\phi^b\rangle =
         S{\phi_0^2\over a^2\eta^2}.
\end{equation}
with a constant $S$. If this ansatz holds, the ``density'' of
the stiff source is also proportional to $\eta^{-2}$ and the
dynamics are scale-invariant with respect to the horizon size
(or Hubble radius).  We shall see later that this choice is
self-consistent.

Before the phase transition (at temperatures $T\gtrsim\phi_0$),
the potential (or lagrange multiplier) term in the lagrangian is
irrelevant and we expect $\phi^2\sim T^2$ due to thermal
fluctations. After the phase transition, the field is pinned to
the vacuum manifold.  In the large-$N$ limit, the
distribution of the individual components $\phi^a$ becomes a
Gaussian peaked around $\phi^a=0$ with a variance given by
$\langle\phi^2\rangle=\phi_0^2$ (see Appendix).  Therefore, we
simply Fourier transform,
\begin{equation}
\phi^a({\bf x},\eta) =
{\phi_0\over\sqrt{4\pi N}}
\int d^3k\,\alpha^a_{\bf k}\phi_k(\eta)e^{i{\bf k}\cdot{\bf x}}
\end{equation}
where the prefactor enforces the constraint $\phi^2 =\phi^2_0$
and the $\alpha_{\bf k}^a$ are {\em Gaussian} random variables of
mean $0$ and normalized to unit variance for later ease of
calculation.  They are uncorrelated for unequal $a$ and ${\bf k}$;
for example,
\begin{equation}
\langle\alpha_{\bf k}^a\alpha_{\bf q}^b\rangle
  = \delta^{ab}\delta^3({\bf k}+{\bf q}),
\end{equation}
where $\delta^{ab}$ and $\delta^3({\bf k})$ are Kronecker and Dirac
deltas, respectively,
and higher (even) order correlations are sums of appropriate products of
this two-point function; averages of odd numbers of the
$\alpha_{\bf k}^a$ vanish.

In a universe with scale factor $a\propto\eta^\alpha$ and the
scaling ansatz, Eq.\ (\ref{sigmaeom}) becomes the {\em linear}
equation
\begin{equation}
\phi_k'' + {2\alpha\over\eta}\phi_k' + k^2\phi_k = -{S\over\eta^2}\phi_k.
\end{equation}
This has solution
\begin{equation}\label{phik}
\phi_k(\eta)\equiv k^{-3/2} f(k\eta);\qquad
f(x) = {1\over{\cal N}_0^{1/2}}x^{1/2-\alpha}J_{1+\alpha}(x)
\end{equation}
where the order of the Bessel function (and the value of the
constant $S$) has been chosen so that the solution has a white-noise
power spectrum on superhorizon scales, and ${\cal N}_0$ is a
normalization constant. The presence of the simple form
$f(k\eta)$ indicates the expected scaling nature of the
solution: $k\eta$ is the ratio of the horizon
($\eta\propto1/{\cal H}$) to the length scale ($k\sim1/\lambda$).

Given this solution, we can calculate the stress energy
$\Theta_{\mu\nu}$ required for the formalism of Section
\ref{sec:formal}. Most generally,
\begin{equation}
\Theta_{\mu\nu} = {2\over\sqrt{-g}}\left[
{\partial(\sqrt{-g}{\cal L})\over\partial g^{\mu\nu}} -
  {\partial\over\partial x^\alpha}
{\partial(\sqrt{-g}{\cal L})\over\partial (\partial_\alpha g^{\mu\nu})}\right].
\end{equation}
In this case, the second term vanishes and the ``potential'' (lagrange
multiplier) term is zero when the vacuum manifold conditions are
satisfied. So,
\begin{equation}
 \Theta_{\mu\nu} = \partial_\mu\phi^a \partial_\nu\phi^a -  {1\over 2}
 \eta_{\mu\nu}\eta^{\alpha\beta}\partial_\alpha\phi^a \partial_\beta\phi^a.
\end{equation}
Particular components are
\begin{eqnarray}
\Theta_{00} &=& {1\over 2}\left(\phi^{a\prime}\right)^2 +
              {1\over 2}\left(\nabla\phi^a\right)^2  \\
\Theta_{0i} &=& \phi^{a\prime}\nabla\phi^a \label{Theta0i}\\
\Theta_{ij} &=& \partial_i\phi^a\partial_j\phi^a + {1\over 2}\delta_{ij}
 \left[\left(\phi^{a\prime}\right)^2-\left(\nabla\phi^a\right)^2\right].
\end{eqnarray}
The solutions for the $\phi^a$ can be inserted into these expressions to
calculate the components in this model. We find that
\begin{equation}
\langle\Theta_{00}\rangle \propto {1\over\eta^2}
\end{equation}
in a matter-dominated universe, so the scaling solution holds. In a
radiation-dominated universe, there is also a factor of
$\ln(\eta/\eta_1)$, so the scaling is modified by a logarithmic term. We
also find that the ``pressure,'' $\langle\Theta\rangle$ vanishes in a
matter-dominated universe and is $(1/3)\langle\Theta_{00}\rangle$
during radiation-domination. We shall calculate $\Theta_{0i}$ below when
considering density fluctuations.

\subsection{Density Perturbations and Power Spectra}

When the sigma-model solution, Eq.\ (\ref{phik}), and the
fourier transform of the expression for the stress-energy from
Eq.\ (\ref{Theta0i}) are inserted into Eq.\ (\ref{md_mode})
above for the linear perturbation to the matter density, the
result is
\begin{equation}\label{deltak}
\delta({\bf k}) =
{4\pi G\over10}\eta^2\int d\eta\,\partial_i\Theta_{0i}({\bf k})
        = {G \phi_0^2\over10N}\eta^2
  \int d^3q\, {\bf q}\cdot{\bf k} \; \alpha^a_{{\bf k}-{\bf q}}\alpha^a_{\bf q}
       \int d\eta \, \phi'_{|{\bf k}-{\bf q}|}(\eta)\phi_q(\eta)
\end{equation}
where we have assumed the power law exponent $\gamma=0$, or
$\int d\eta\,\Theta_{0i}\approx{\rm const}$ in Eq.\ (\ref{md_mode}),
to be justified below.  This form enables the computation of
$n$-point power spectra.

In general, a function of the form
$\langle\delta({\bf k})^n\rangle$ contains the product of $2n$
Gaussian random variables $\alpha_{\bf k}^a$; thus even though the
odd moments of the field $\phi$ may vanish, the corresponding
moments of the density field may be nonzero, and thus may be
non-Gaussian even to linear order.

In addition, many of the terms in the expansion of the order $n$
moment of the $\alpha_{\bf k}^a$ will vanish due to the form of the
integrand above. For example, the power spectrum
$\langle\delta({\bf k})\delta({\bf k}')\rangle$ contains a term with
\begin{equation}
{\bf k}\cdot{\bf q}\,{\bf k}'\cdot{\bf q}'\,
\delta^{aa}\delta^{a'a'}\delta^3({\bf k})\delta^3({\bf k}')N^{-2},
\end{equation}
where the Kronecker and Dirac deltas result from the mean of
four of the $\alpha^a_{\bf k}$ from the previous equation. This
term vanishes due to the presence of ${\bf k}'\delta^3({\bf k}')=0$.
Similarly, in any moment, any term which contains a combination
of Kronecker deltas that do {\it not} contract overall to $N$
will also contain a factor like ${\bf k}'\delta^3({\bf k}')$ and will
thus integrate to zero (this factor is effectively
$\langle\delta_{\bf k}\rangle=0$); the remaining terms behave like
\begin{equation}
 {\bf k}\cdot{\bf q}\,{\bf k}'\cdot{\bf q}'\,
 \delta^{a'a}\delta^{a'a}\delta^3({\bf k}+
 {\bf q}')\delta^3({\bf k}'+{\bf q})N^{-2}
\end{equation}
in the calculation of the (2-point) power spectrum. In this
case, the Kronecker deltas contract to one power of $N$ and the
Dirac deltas are equivalent to $\delta^3({\bf k}+{\bf k}')$ after the
integrals have been performed. For an $n$-point spectrum this
works out to $N\times N^{-n}=1/N^{n-1}$, where the first factor
$N$ comes from the contraction of the Kronecker deltas and
$N^{-n}$ from the $n$ prefactors of $\delta({\bf k})$ in Eq.\
(\ref{deltak}).  Moreover, when all the remaining integrals have
been completed, there will still be a delta-function to enforce
the requirement that the sum of the ${\bf k}_i$ be zero as
expected. Therefore, the moment will behave like
\begin{equation}
\langle\delta({\bf k}_1)\cdots\delta({\bf k}_n)\rangle \propto
\left({G\phi_0^2\over10}\eta^2\right)^n {1\over N^{n-1}}
   \delta^3\left(\sum{\bf k}_i\right).
\end{equation}

In fact, we can proceed further without actually doing any
calculating.  In addition to the factors in the previous
equation, the $n$th moment will contain an integral over $d^3q$
of $n$ quadratics in ${\bf q}$ and the ${\bf k}_i$ as well as a
product of $n$ integrals of the form
$\int d\eta\,\phi'_p\phi_{p'}$, where $p,p'$ are the lengths of linear
combinations of ${\bf q}$ and the ${\bf k}_i$. As in Davis {\it
et al.}\cite{davis}, this integral can in general be written as
a function of the configuration and overall scale ({\it i.e.},
$k\equiv k_1$) of the ${\bf k}_i$ by writing ${\bf q}=k{\bf u}$
and $\eta=s/k$, and all of the remaining ${\bf k}_i=k ({\bf
k}_i/k)$.  Furthermore, we use the scaling of the $\phi$ fields,
\begin{equation}
 \phi_p(\eta) = p^{-3/2}f(p\eta),\qquad \phi'_p(\eta)=p^{-1/2}f'(p\eta)
\end{equation}
in the $\eta$ integral.  We can then pull out all of the factors of
scale to be left with $k^{3-n}$ in front. Thus, the $n$th
moment (factoring out the momentum-conserving $\delta$ function) is
given by
\begin{equation}\label{Pn}
P({\bf k}_1,\ldots,{\bf k}_{n-1}) =
\left({G\phi_0^2\over10}\eta^2\right)^n {k^{3-n}\over N^{n-1}}
         g_n(k\eta, \hbox{configuration}),
\end{equation}
where ``configuration'' refers to the shape (but not the scale)
of the $n$-sided polyhedron defined by the ${\bf k}_i$. (For
$n\ge4$, the irreducible moment $P_n$ differs from the reducible
$\langle\delta_1\cdots\delta_n\rangle$ by terms that vanish
unless two of the wavevectors satisfy ${\bf k}_i+{\bf k}_j=0$;
for continuous ${\bf k}$ this is a set of measure zero, and in
any case does not apply to the configurations usually examined
when considering data.) For suitably late times and subhorizon
scales ($k\eta\rightarrow\infty$), the function $g_n$ should go
to a constant, dependent only on the configuration of the
figure.

In particular we find that the subhorizon power spectrum is
Harrison-Zel'dovich, $P(k)\propto k$, the bispectrum $B({\bf k}_1, {\bf k}_2)$
is independent of scale, and all higher moments go as a negative power
of $k$; that is, they decrease with decreasing scale. However, this
calculation is only valid for scales which have entered the horizon in
the matter-dominated era. Thus, the spectrum of perturbations on scales
smaller than the horizon at matter-radiation equality cannot be
calculated from these formulae.

Thus far, we have neglected the fact that we must take the
large-$N$ limit of these quantities. If we normalize them to the
(two-point) power spectrum $P(k)$ through a quantity such as
$J_3$ (the volume integral of the two-point correlation
function) or $\sigma_8$, we get a finite numerical value for the
combination
$G\phi_0^2/\sqrt{N}\approx3\times10^{-5}\sigma_8$\cite{davis}
(where $\sigma_8^2$ is the variance of the mass distribution in
spheres of $8\,h^{-1}{\rm Mpc}$; the variance in the number density of
galaxies on that scale is measured to be $\sigma_8({\rm gal})=1$
\cite{sig8}). The scaling of this value, $\phi_0^2\propto\sqrt{N}$ is a
consequence of the analytic calculations reproduced in this
section for large $N$. However, the numerical value comes from
various simulations. Comparison of the normalization (to
$\sigma_8$) for specific values of $N\le6$ in simulations (which
explicitly include topological-defect field configurations not
present for large $N$), indicates that this value for $\phi_0$
may be as much as a factor of 10 too high. Although the
calculations reproduced in this section show that we should
expect $\phi_0^2\propto\sqrt{N}$ in the large-$N$ limit, for
these low values of $N$ that have been numerically investigated,
such scaling has not yet been reached
\cite{spergelprivate}.  In any case, none of these values are
expected to be known to better than about 50\%\cite{penetal}
(The low-$N$ cases are in the process of being solved by using
exact Green's functions rather than direct integration of the
equations of motion presented here. This is expected to give
more accurate results
\cite{turokprivate}.)

Thus, the $n$-point spectra work out to be
\begin{equation}
P_n = \left({G\phi_0^2\over10\sqrt{N}}\eta^2\right)^n g_n
{k^{3-n}\over N^{n/2-1}}.
\end{equation}
In the large-$N$ limit, these quantities decrease with $N$ for
all $n\ge3$.  However, the leading term in $1/N$ causes a
non-Gaussian distribution to this order, and we expect these
calculations to be at least approximately valid for $N$ greater
than a few, since the character of the sigma-field is
approximately the same in the absence of topological defects
such as strings ($N=2$) or textures ($N=4$).


The asymptotic value of $g_2$ for large $k$ (small scales) is
given by
\begin{equation}
  g_2 = \int d^3u \,
   \left[{\bf u}\cdot{\bf\hat k}\left(2{\bf u}\cdot{\bf\hat k}-1\right)\right]
                        I(|{\bf\hat k}-{\bf u}|, u; k\eta)
\end{equation}
as in \cite{davis}, where
\begin{equation}
I(a,b;x) = \int^x ds\, {f(as)f'(bs)\over a^{3/2}b^{1/2}}.
\end{equation}
We have already assumed that $k\eta\gg1$ so
we may integrate by parts and eliminate surface terms to
consolidate the various integrals $I(a,b;k\eta)\equiv I(a,b)$ that appear
(see below and Fig.\ (\ref{fig:gn})).
The corresponding value for the bispectrum, $g_3$, is
\begin{eqnarray}
 g_3({\bf v}) = \int  d^3u \,
     {\bf u}\cdot{\bf\hat k} \, I(|{\bf\hat k}-{\bf u}|, u)
\bigg[
 (&&1+2{\bf v}\cdot{\bf\hat k}-2{\bf u}\cdot{\bf\hat k}+
      v^2-2{\bf u}\cdot{\bf v})(2{\bf u}\cdot{\bf v}-
      2{\bf v}\cdot{\bf\hat k}-v^2)\times \nonumber \\
  && I(|{\bf\hat k}+{\bf v}-{\bf u}|,|{\bf u}-{\bf\hat k}|)
     I(u,|{\bf\hat k}+{\bf v}-{\bf u}|)+ \nonumber\\
 (v^2+2{\bf u}\cdot{\bf v})(2{\bf u}\cdot{\bf\hat k}+
         2{\bf u}\cdot{\bf v}&&+v^2-1)
         I(u,|{\bf u}+{\bf v}|)I(|{\bf u}-{\bf\hat k}|,|{\bf v}+{\bf u}|)
\bigg]
\end{eqnarray}
where ${\bf v}={\bf k}_2/k_1$ embodies the dependence upon the
configuration of the wavevectors. For an equilateral triangle,
$k_2=k_1=k$ and ${\bf u}\cdot{\bf v}=-1/2$.

Note that these quantities do have a residual scale dependence
in the form of the upper limit of the innermost integrals $I$.
In Fig.\ (\ref{fig:gn}), we plot numerical integrations of $g_2$
and $g_3$ as a function of $k$ where we integrate from horizon
crossing ($k\eta=1$) to the present epoch ($k\eta=k\eta_0$) at
each wavenumber. Both of these quantities reach their asymptotic
values ($g_2\approx12, g_3\approx1.6$) by
$k^{-1}\approx10^3\,h^{-1}{\rm Mpc}$, well outside of the range of our
ability to reliably measure higher-order correlation functions.
This quick approach to constant values indicates that, as
suspected, $\int d\eta\,\partial_i\Theta_{0i}\approx{\rm const}$,
so we are justified in assuming the power-law exponent
$\gamma\approx0$ in Eqs.\ (\ref{md_mode}) and (\ref{rd_mode})
above---the situation mimics a simple growing mode. Physically,
this is simple to understand: as a given mode enters the
horizon, it comes into causal contact fairly quickly, on the
order of a Hubble time.  After that, the field is roughly
static, so its momentum density $\Theta_{0i}$ is small, and the
matter perturbations are only generated on scales near the horizon.
Subsequently, they evolve under the influence of small-scale
Newtonian gravity alone.  (However, it is not hard to imagine
some other kind of stiff source with ``ordering physics'' that
continues to be active on smaller scales which would result in
density fluctuations that do not mimic the Newtonian
growing-mode result.)

One traditional way to characterize the bispectrum is through
the ratio
\begin{equation}\label{qk}
Q\equiv {B({\bf k}_1, {\bf k}_2)\over P_1P_2+P_1P_3+P_2P_3}
\end{equation}
which for an equilateral triangle becomes $Q=B(k)/3P(k)^2$. The
usual hierarchical result from the Newtonian analysis with
Gaussian initial conditions is $Q={\rm const}$ from second-order
perturbation theory. With the
non-Gaussian initial conditions from the sigma model, we instead
have $Q\propto k^{-2}$ from {\em linear} theory.

The question remains, then, on what scales do we expect these
linearly evolved moments to dominate over higher-order effects?
{}From Section \ref{sec:nlstiff} above, the leading nonlinear
contribution to the higher moments is
\begin{equation}
 P_{n, \rm nl}(k)=Q_nP_2^{n-1}=
     Q_n\left({G\phi_0^2\over10\sqrt{N}}\eta^2\right)^{2(n-1)}(g_2k)^{n-1},
\end{equation}
where this equation is actually only correct for $n=3$ with
equilateral triangles, where $Q_3=3Q$. (For $n>3$, the
right-hand side of this equation should be a sum over the
different ``tree-level'' graphs connecting the labelings of the
wavevectors, each term with a corresponding constant $Q_{n;a}$
\cite{fry84,bernardeau,baumgart+fry}. However, for analyzing the
scale-dependencies here, the form above shall suffice.)

Consider the quantities $d_n = P_n/P_2^{n/2}$ which normalize the
$n$-point spectra to $|\delta({\bf k})|^n$. We see that the linear
contribution is, of course, constant,
\begin{equation}
d_{n,\rm lin} = {g_n\over g_2^{n/2}} k^{3(1-n/2)}N^{1-n/2}
\end{equation}
whereas the nonlinear contribution grows with time,
\begin{equation}
d_{n,\rm nl} =
   Q_n\left({G\phi_0^2\over10\sqrt{N}}g_2k\eta^2\right)^{n/2-1}
\end{equation}
At a given wavenumber, the
nonlinear contribution grows with time with respect to the linear
contribution.  A similar result for spatial correlation functions was
derived for $n=3$ in \cite{fry+scherrer}, where they also included the
effects of a primordial four-point function (whose contribution to the
skewness also grows with time).

The ratio of the nonlinear contribution to that of the linear
evolution is
\begin{equation}
{P_{n, \rm nl}\over P_{n, \rm lin}} = Q_n
\left[{G\phi_0^2\over10\sqrt{N}}\right]^{n-2} N^{n/2-1}
{g_2^{n-1}\over g_n}(k\eta)^{2n-4}
\end{equation}
which goes as a positive power of $k$ for $n>2$, so the nonlinear terms
become more important on small scales, as expected. This is a function
of $k\eta$, the ratio of the length involved to the horizon size, so the
scaling nature of the solution is preserved.

The scale of this crossover from the domination of the linear to the
nonlinear evolution of the fields depends on the values of the $g_n$ for
a given configuration of wavevectors and $N$ (although the powers of $N$
explicitly cancel in the above expression, recall that the normalization
of $\phi_0$ is such that factor in brackets above is
constant). Crossover thus occurs at scale
\begin{equation}\label{knl}
k_{\rm nl}^{-1}=\eta\left(Q_n{g_2^{n-1}\over g_n}\right)^{1/(2n-4)}
N^{1/4}\left(G\phi_0^2\over10\sqrt{N}\right)^{1/2}.
\end{equation}

Actually, the full evolution of the three-point function depends
on the four-point function (or trispectrum $T({\bf k}_1, {\bf
k}_2, {\bf k}_3)$) as well\cite{fry+scherrer}, which we do not
explicitly calculate here, but we expect this crossover scale to
be independent of it, since from above we have $T\propto k^{-1}$
so the trispectrum falls on small scales, at least for those
greater than that of matter-radiation equality. Moreover, the
value of $n$ only enters these equations through the prefactor
involving the $Q_n$ and $g_n$, which are not expected to vary
greatly with $n$. So, for increasing $n$, the $n$-dependent part
of the first factor in parentheses will be come less important,
and the crossover scale becomes approximately
\begin{equation}
k_{\rm nl}^{-1}\approx\eta g_2^{1/2} N^{1/4}
\left(G\phi_0^2\over10\sqrt{N}\right)^{1/2};\qquad({\rm large\ } n).
\end{equation}

If we naively plug the asymptotic values of $g_2\approx12$ and
$g_3\approx1.6$ into Eq.\ (\ref{knl}), with the hierarchical value
for equilateral triangles of $Q_3 = 3Q = 34/7$ we find
\begin{equation}
   k_{\rm nl}^{-1} \approx 219 N^{1/4} \sigma_8^{1/2} h^{-1}{\rm Mpc};
     \qquad(n=3)
\end{equation}
for $G\phi_0^2/\sqrt{N}\approx3\times10^{-5}\sigma_8$. If we let this
quantity be a factor of 10 smaller as indicated above, then the
crossover scale is reduced to $k_{\rm nl}^{-1} \approx69 N^{1/4}
\sigma_8^{1/2}h^{-1}{\rm Mpc}$. To calculate the scale at which the primordial
four-point function or trispectrum becomes significant, we will use
the hierarchical pattern $T\approx16RP^3$ for tetrahedral
configurations, with $R\sim Q^2$,\cite{fry84} so $Q_4\approx16Q^2\sim5$
(although observations\cite{baumgart+fry} are typically
$R\approx1$.) Thus,
\begin{equation}
k_{\rm nl}^{-1}\approx
(17-54) g_4^{-1/4} N^{1/4}\sigma_8^{1/2}h^{-1}{\rm Mpc};
   \qquad(n=4)
\end{equation}
spanning the range of possible normalizations of $\phi_0$, where
we have left the integral $g_4$ unevaluated, but expect it to be
of order unity.

Even this scale, unfortunately, is just beyond the
largest considered in surveys for higher-order power spectra,
which have gone out to $k^{-1}\approx20\,h^{-1}{\rm Mpc}$
\cite{baumgart+fry}.  We expect only the {\em nonlinear}
contribution to the three-point function to be important on the
smallest observable scales. If instead we use the second,
large-$n$ form of $k_{\rm nl}$, we find
\begin{equation}
   k_{\rm nl}^{-1}\sim(11-36)N^{1/4}\sigma_8^{1/2}h^{-1}{\rm Mpc};
   \qquad({\rm large\ } n)
\end{equation}
This is more possibly in the range of observed and observable
scales, but still (for $N\gtrsim6$ where we expect to believe these
results) barely on the edge of current observations of the
bispectrum and trispectrum.

Thus, the usual Newtonian analysis
of higher-order correlation should suffice there.
Unfortunately, that means that it will be more difficult to
distinguish the nonlinear sigma model from simple Gaussian
theories on small scales using the mass distribution alone.
However, as discussed in \cite{penetal}, the large-scale
normalization of the nonlinear sigma model to the COBE results
may require an unreasonably large bias in order to match the
galaxy distribution.

More problematic is the calculation on scales that entered the
horizon during the radiation-dominated period. A complete
calculation requires the numerical integration of the equations
of motion of the sigma model fields and the radiation and matter
fluids as in \cite{penetal}.  They present a possible
parametrization of the transfer function which takes the power
spectrum from $P(k)\propto k$ on large scales to its small-scale
form (usually $P(k)\propto k^{-3}$, possibly modulated by the
logarithmic growth of modes in the radiation-dominated
universe). However, the equivalent form has not been calculated
for the three-point function. However, we do not expect the
linear evolution to dominate on small scales. Higher order or
fully nonlinear effects tend to dominate on the smallest scales.
In order for the linear correlations to be observable below the
scale of matter-radiation equality, the linear bispectrum would
have to increase relative to the nonlinear contribution on these
scales.  Moreover, because of the form of Eqs.\ (\ref{nsource}),
the relative importance of higher order effects with respect to
linear effects is again expected to be the same as in the purely
Newtonian case.

\subsection{Comparison with Observations}

As we have seen, the nonlinear sigma model is not expected to
give non-Gaussian results for higher-order correlation functions
on as-yet observed scales. So far, all observations are
consistent with Newtonian hierarchical results with Gaussian
initial conditions. Baumgart and Fry \cite{baumgart+fry}
actually analyze the $n$-point power spectra for two galaxy
surveys and find that $Q(k)=B/3P^2\simeq4/7$ (for equilateral
triangles) for $k^{-1}\lesssim 10\,h^{-1}{\rm Mpc}$. This is exactly
the value expected from the Newtonian theory with Gaussian
initial conditions. They also calculate $R(k)=T/16P^3\approx1$
(for regular tetrahedral configurations), a result which is
weakly consistent with the hierarchical expectation.

More recent results of analyses of the IRAS dataset are also
consistent with the hierarchical expectations. Bouchet {\it et al.}
\cite{bouchetetal} have calculated the volume-averaged $n$-point
functions $\bar\xi_n$ for $n\le5$ and again find that the
hierarchical form holds: $\bar\xi_n\propto\bar\xi_2^{n-1}$.
Moreover, Nusser {\it et al.}\cite{nusseretal} have reconstructed the
{\em initial} one-point probability density function for the
smoothed IRAS density field (using an algorithm based on
Newtonian gravity and the Zel'dovich approximation) and found it
to be consistent with a Gaussian distribution on scales smaller
than about $70\,h^{-1}{\rm Mpc}$.

It may be possible to examine the $n$-point functions at still
larger scales using two-dimensional angular correlations
\cite{fry+seldner}. However, an angular analysis introduces
errors due to uncertainties in the luminosity function of
galaxies and consequently the survey's selection function,
especially on the large scales we are most interested in.
Nonetheless, Gazta\~naga\cite{gaz} has estimated the $n$-point
angular correlation functions for the APM dataset and found that
the results are consistent with an initially Gaussian
distribution for scales out to about $30 h^{-1}{\rm Mpc}$; however,
this is still below the scales required to see any possible
effects of a sigma-model field.

Precise measurements of the bispectrum and trispectrum out to
scales of $k^{-1}\approx100\,h^{-1}{\rm Mpc}$ should be sufficient to
begin to more explicitly test this model. We should expect the
next generation of survey results for higher-order correlations
to at least begin to probe the scales on which primordial
non-Gaussianity from a sigma-model source should be observable,
at least for moderate values of $N$. That we do not yet observe
a marked increase in the three-, four-, or five-point functions
(or their spectral counterparts) on the largest scales implies
that the density field on those scales remains dominated by
Newtonian evolution. If this observational trend continues, then
any crossover to linear evolution will be on still larger
scales, and we will be able to make stronger statements about
the normalization or existence of any sigma-model field ({\it e.g.},
using Eq.\ (\ref{knl})).

We re-emphasize the fact that observations of the scaling
hierarchy of correlation function do not imply Gaussian initial
conditions, even on the scale of the observations. As we have
seen here, the correlation functions on small scales are
dominated by the contributions from the initially Gaussian
component of the density field, which evolves nonlinearly into
the quasi-Gaussian scaling hierarchy. This behavior is expected
to be generic to any scenario with non-Gaussian initial
conditions, as the nonlinear evolution should in any case
dominate on small scales.

Other observational ramifications of the sigma-model may be more
damning even today. As Pen {\it et al.}\cite{penetal} point out,
normalization of the microwave distortions produced in these
models to the COBE observations require what may be an
uncomfortably high bias, $b=1/\sigma_8\approx2/h$. Moreover, the
detailed fit of the power spectrum of mass fluctuations
(determined from numerical simulations) to the measured QDOT
galaxy distribution seems to require a bias of order 6 on scales
of $20\,h^{-1}{\rm Mpc}$.

\section{Conclusion}
\label{sec:conclusion}
We have developed a formalism to examine the quasi-nonlinear
behavior of perturbations in a universe with a ``stiff'' source.
In particular, we have shown that their evolution is
formally very similar to those encountered in a purely Newtonian
analysis with an initial perturbation spectrum, but without a
source. In the particular case of the nonlinear sigma model, we
find that the behavior of higher correlation functions of the
density field on extreme subhorizon scales is exactly the same
as that which occurs with primordial {\em adiabatic}
perturbations. On larger, currently unobservable, scales, two
effects occur that might differentiate the sigma model. First,
the sigma model (at least at moderate $N$), has an initially
non-Gaussian distribution, which might be observed on scales of
several hundred megaparsecs. Furthermore, we find that the
density perturbations are actually being created on scales
comparable to the horizon, where the self-ordering physics of
the source is occurring. Unfortunately, the prospects of observing
effects on these scales are slim. However, it may be possible to
invent models in which the equivalent self-ordering occurs on
smaller scales as well. Unfortunately, the chief
candidates---defect theories like cosmic strings---are
ill-suited for this analysis, because the gravitational
effects of defects generally occur on too small a scale.
Therefore, a quasi-nonlinear analysis is insufficient and
full-blown numerical simulations must be performed.

There are several possible extensions to the work we
have presented here. We have concentrated on scalar
perturbations, because those result in the readily-observable
density fluctuations, even when higher-order terms in the matter
variables are present. By considering vector and tensor
perturbations, we can go beyond this and calculate quantities
like the full velocity field (including any possible vortical
part), as well as the tensor perturbations (gravitational
radiation) from the stiff source.

If we consider a universe filled with both matter and radiation
fluids, it should be possible to use this formalism to make
accurate calculations of microwave anisotropies in these
theories. If we have also calculated the gravitational
radiation, then we can separate out the scalar and tensor
components \cite{tensorcmb}.

Finally, it should be possible to extend the cosmological
post-Newtonian suite of approximations\cite{postnewton} to
include the possibility of a stiff source. Because this involves
going to a higher order in the metric perturbation, things
become considerable more complicated, but such approximations
are often useful even when perturbations have become extremely
nonlinear.

\acknowledgements
Presented as a dissertation to the Department of Astronomy and
Astrophysics, The University of Chicago, in partial fulfillment
of the requirements for the Ph.D. degree. I would like to thank
Albert Stebbins for many useful discussions and especially thank
my advisor, Joshua Frieman, for guidance throughout the project.
This work was supported in part by the DoE at the University of
Chicago and by the DoE and by NASA through grant NAGW 2381 at
Fermilab.

\appendix
\section*{Distribution of the Fields in the $O(N)$ model}
\label{app:Gaussian}

In the nonlinear $O(N)$ sigma model, the fields $\phi^a$ are
constrained to lie upon the vacuum manifold, an $(N-1)$-sphere
of radius $\phi_0$, where they will be distributed uniformly in
the absence of any further symmetry breaking. Thus, the
probablity density will be proportional to the solid angle on
the $(N-1)$-sphere.

We choose $N-1$ polar coordinates ($\varphi$, $\theta_1$, $\ldots$,
$\theta_{N-2}$) such that the solid angle is
\begin{equation}
d\Omega_N = d\varphi \, \sin\theta_1 \,d\theta_1 \cdots \, \sin^i\theta_i
\,d\theta_i \cdots \,\sin^{N-2}\theta_{N-2}\,d\theta_{N-2}.
\end{equation}
Thus, the distribution of fields in polar coordinates is given by
\begin{equation}
p(\varphi,\{\theta_i\})\,d\varphi\,d\theta_1\cdots\,d\theta_{N-2}
            \propto d\Omega_N.
\end{equation}

We are concerned, however, with the distribution of the $\phi^a$, the
Cartesian components of the fields. In this coordinate system, there is
always one component given by
$z \equiv\phi^a/\phi_0 = \cos\theta_{N-2}$, and, due to the
symmetry of the system, we can calculate the distribution of any single
Cartesian component:
\begin{eqnarray}
p(z)&=&\int p(\varphi, \{\theta_i\})
    \,d\varphi\,d\theta_1\cdots\,d\theta_{N-2}\,
             \delta(z-\cos\theta_{N-2})\nonumber\\
    &=& A' \int d\Omega_N \delta(z-\cos\theta_{N-2})\nonumber\\
    &=& A' \int d\varphi \, \int d\theta_1\,\sin\theta_1 \, \cdots
   \int d\theta_{N-3}\,\sin^{N-3}\theta_{N-3}\int d\theta_{N-2}\,\sin^{N-2}
             \delta(z-\cos\theta_{N-2})\nonumber\\
    &=& 2\pi A' \int d\theta_1\,\sin\theta_1 \, \cdots
                \int d\theta_{N-3}\,\sin^{N-3} \theta_{N-3}
	        \int_{-1}^1 dy\,(1-y^2)^{(N-3)/2} \delta(z-y)\nonumber\\
    &=& A (1-z^2)^{(N-3)/2}
\end{eqnarray}
where $A, A'$ are constants determined by the requirement that
$\int dz\,p(z) = 1$, and we have made the change of variables
$y=\cos\theta_{N-2}$.  This gives
\begin{equation}
p(z)\,dz = {1\over\sqrt\pi}{\Gamma(N/2)\over\Gamma(N/2-1/2)}
\left(1-z^2\right)^{(N-3)/2} \,dz.
\end{equation}
This distribution has mean $\langle z\rangle=0$ and variance
$\langle z^2\rangle=1/N$ (which can be calculated directly or
seen from the constraint $\sum\phi^a\phi^a=\phi_0^2$). The
higher moments are given by
\begin{equation}
\langle{z^m}\rangle=
{1\over\sqrt\pi}{\Gamma(N/2)\Gamma(m/2+1/2)\over\Gamma(N/2+m/2)}
\qquad (m \hbox{ even}),
\end{equation}
which behave as $O(1/N^{m/2})$ for large $N$;
the odd-$m$ moments vanish due to the symmetry of the distribution.
For large $N$,
\begin{equation}
p(z) \rightarrow \sqrt{N\over2\pi}\left(1-z^2\right)^{N/2}.
\end{equation}

We wish to compare this to a Gaussian distribution with
some width $\sigma$,
\begin{equation}
g(z) = {1\over\sqrt{2\pi\sigma^2}} \exp\left(-z^2/2\sigma^2\right); \qquad
\langle{z^m}\rangle =
{\left(2\sigma^2\right)^{m/2}\over\sqrt\pi} \Gamma(m/2+1/2)
\quad (m \hbox{ even}).
\end{equation}
For small $z^2$ and large $N$,
\begin{equation}
p(z) \rightarrow \sqrt{N\over2\pi}\left(1-{Nz^2\over2}\right),\qquad
g(z) \rightarrow {1\over\sqrt{2\pi\sigma^2}}\left(1-{z^2\over2\sigma^2}\right),
\end{equation}
which implies $\sigma^2=1/N$ to first order in $1/N$, as expected from
$\langle{z^2}\rangle=1/N$ and the expression for the higher moments of the
Gaussian distribution at large $N$.
For $z\approx1$,
both $p(z), g(z)\rightarrow0$ for large $N$ or small $\sigma$ (although
the ratio $p(z)/g(z)$ can be quite large for a finite value of $N$; this
is because the Gaussian is normalized over the interval
$(-\infty,+\infty)$). Moreover, in the limit of $N\rightarrow\infty$,
both $p(z)$ and $g(z)$ approach the Dirac $\delta$-function. In
Figure \ref{fig:dist} we show $p(z)$ and $g(z)$ for various
values of $N$. Note that for $N\lesssim5$, the departures from
Gaussianity are significant.

Thus, the distribution of the Cartesian components of the field
approaches that of a Gaussian with $\sigma^2 = 1/N$.  For large
$N$, then, the components of the field become more and more
strongly peaked about $z=0$. Note that this distribution is
applicable only on superhorizon scales, where the ordering
dynamics of the field are unimportant. Within the horizon, the
fields will continually organize in order to minimize their
gradient energy as new scales enter the causally-connected
region.


\begin{figure}
\caption{
The integrals $g_2(k)$ (solid) and $g_3(k)$ (dashed) which are
the prefactors in the calcuation of the linear contribution to
the power spectrum $P=P_2\propto g_2 k$ and bispectrum (for
equilateral triangles) $B=P_3\propto g_3 k^0$ respectively (Eq.\
(\protect\ref{Pn})).  We have integrated from horizon-crossing
($k\eta=1$) to the present day ($k\eta=k\eta_0$). In each case
the asymptotic value is reached by $k^{-1}\approx 1000
\,h^{-1}{\rm Mpc}$.  Note that we have already assumed $k\eta\gg1$ in
writing down the expressions for the $g_n$, so the details of
the approach to the asymptotic values ($g_2\rightarrow12.2$ and
$g_3\rightarrow1.6$) should only be taken as indicative.
}
\label{fig:gn}
\end{figure}

\begin{figure}
\caption{
A comparison of $p(z=\phi^a/\phi_0)$
(solid), the actual distribution of fields $\phi^a$ in the
$O(N)$ sigma model, with $g(z)$ (dashed), the corresponding
Gaussian distribution of mean $0$ and variance $1/N$, for $N=3,
5, 10$ and $50$.
}
\label{fig:dist}
\end{figure}

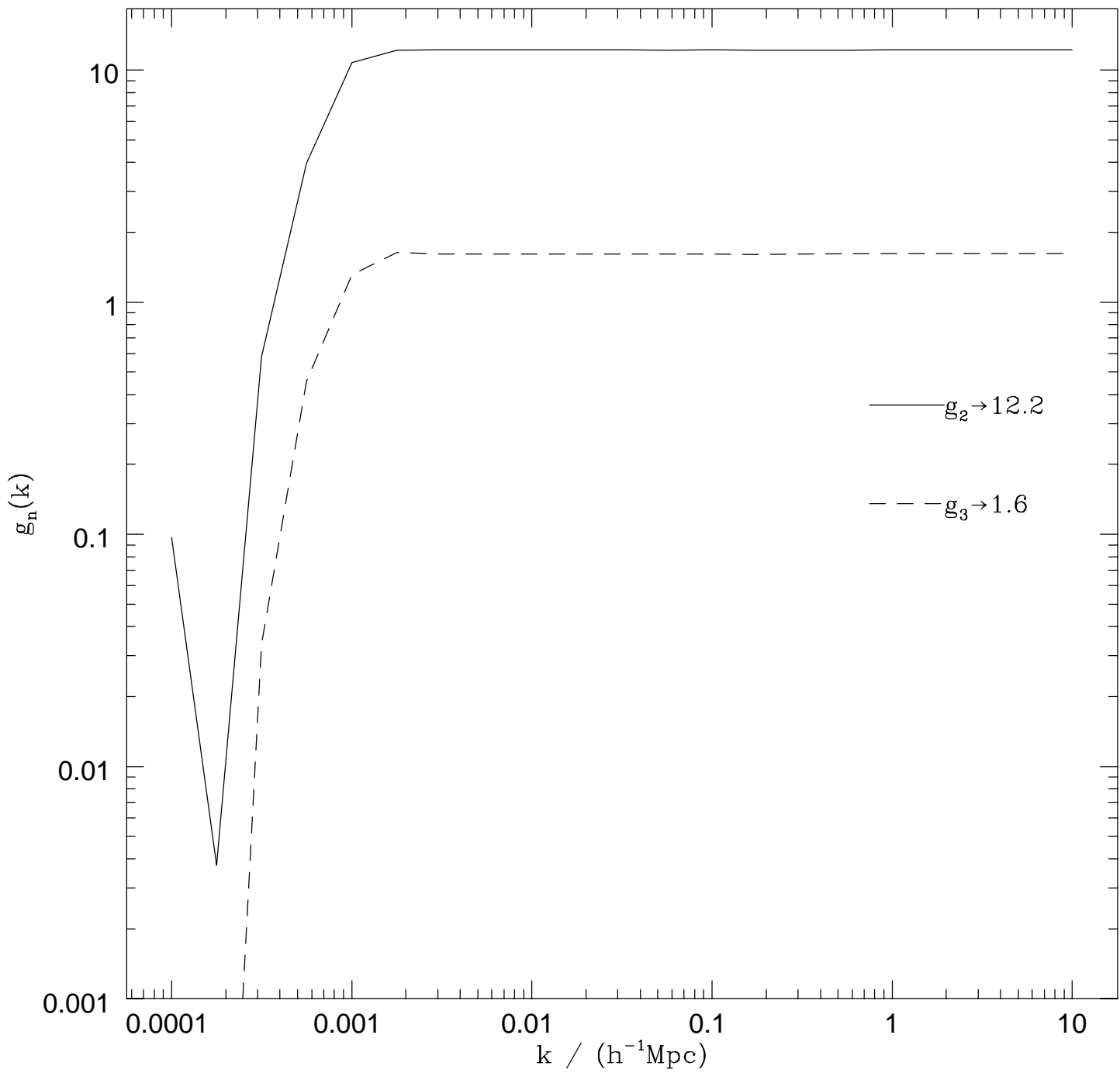
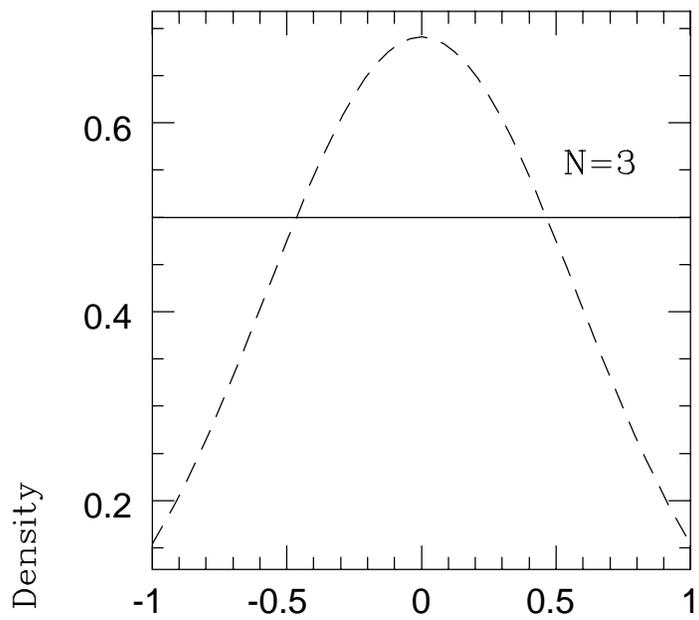
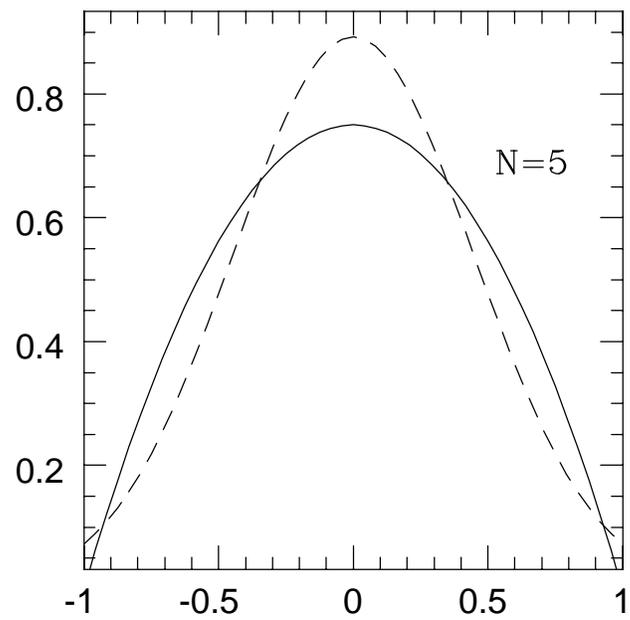
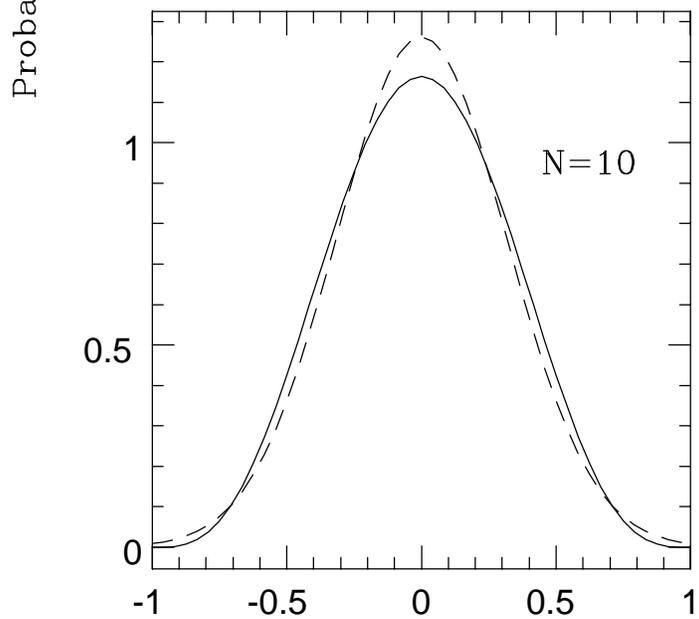
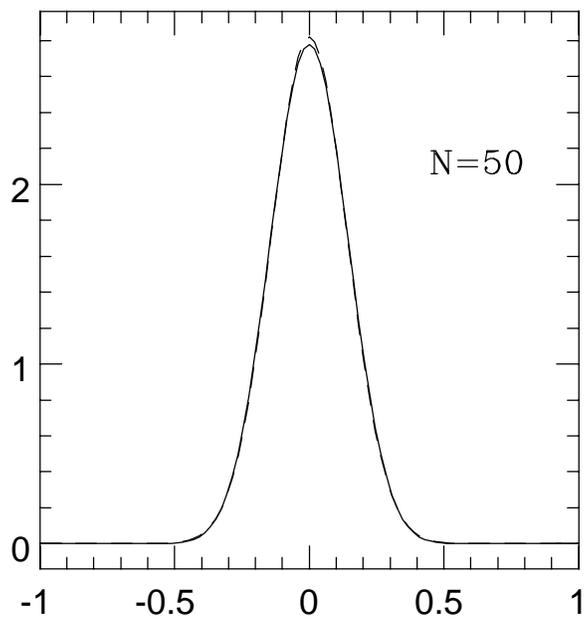
\end{document}